\newif\ifsingle
\singletrue 

\newif\ifFullVersion
\FullVersiontrue 

\ifsingle
\documentclass[11pt,draftclsnofoot, onecolumn]{IEEEtran}		
\else		
\documentclass[10pt,final, twocolumn]{IEEEtran}
\fi

\usepackage{times}
\usepackage{amsmath,bm,dsfont}
\usepackage{mathtools, cuted}
\usepackage{amssymb}
\usepackage{color}
\usepackage{cite}
\usepackage[english]{babel}
\usepackage[bookmarks,colorlinks]{hyperref}

\newtheorem{theorem}{Theorem}
\newtheorem{corollary}{Corollary}

\newtheorem{example}{Example}

\newcommand{\E}{\mathds{E}}
\newcommand{\opt}{^{\rm o}}

\ifsingle
\newcommand{\figWidth}{0.65\columnwidth}

\else
\newcommand{\figWidth}{\columnwidth}

\fi 

\title{\textit{e}Sampling: Energy Harvesting ADCs}

\author{
    \IEEEauthorblockN{Neha Jain, Nir Shlezinger, Bhawna Tiwari,     Yonina C. Eldar, \\
    Anubha Gupta, Vivek Ashok Bohara and Pydi Ganga Bahubalindruni
    } 
    \thanks{
    Parts of this work were accepted for presentation in the European Signal Processing Conference (EUSIPCO) 2020 as the paper \cite{jain_shlezinger_2020}. 
    This project has received funding from the Benoziyo Endowment Fund for the Advancement of Science, the    Estate of Olga Klein – Astrachan, the European Union’s Horizon 2020 research and innovation program under grant No. 646804-ERC-COG-BNYQ, and from the Israel Science Foundation under grant No. 0100101.
    N. Shlezinger and Y. C. Eldar are with the faculty of Mathematics and Computer Science, Weizmann Institute of Science, Rehovot, Israel (e-mail: nirshlezinger1@gmail.com; yonina@weizmann.ac.il).     
    N. Jain, B. Tiwari, A. Gupta, and  V. A. Bohara  are with Dept. of Electronics and Communication Engineering, Indraprastha Institute of Information Technology-Delhi (IIIT-D), New Delhi, India (e-mail: \{nehaj, bhawnat, anubha, vivek.b\}@iiitd.ac.in).      
    P. Ganga is with  Indian Institute of Science Education and Research (IISER) Bhopal, India (e-mail:ganga@iiserb.ac.in).
    }
}

\begin{document}

\maketitle
    \pagestyle{plain}
    \thispagestyle{plain}

\begin{abstract} 
Analog-to-digital converters (ADCs) allow physical signals to be processed using digital hardware. The power consumed in conversion grows with the sampling rate and quantization resolution, imposing a major challenge in power-limited systems. A common ADC architecture is based on sample-and-hold (S/H) circuits, where the analog signal is being tracked only for a fraction of the sampling period. 
In this paper, we propose the concept of \textit{eSampling ADCs}, which extend the structure of S/H ADCs without altering its conversion procedure, while harvesting energy from the analog signal during the time periods where the signal is not being tracked. This harvested energy can be used to supplement the ADC itself, paving the way to the possibility of zero-power consumption and power-saving ADCs. The amount of energy harvested can be increased by reducing the sampling rate. We  analyze the tradeoff between the ability to recover the sampled signal and the energy harvested, and provide guidelines for setting the sampling rate in the light of accuracy and energy constraints.  
Our analysis indicates that \textit{e}Sampling ADCs operating with up to $12$ bits per sample can acquire bandlimited analog signals such that they can be perfectly recovered (up to the distortion induced in quantization) without requiring power from the external source. Furthermore, our theoretical results reveal that \textit{e}Sampling ADCs can in fact save power by harvesting more energy than they consume. Furthermore, we show how these results imply that an \textit{e}Sampling ADC acquiring a bandlimited signal  at Nyquist rate with $8$ bit ADCs can harvest over $15$ dB more energy than it consumes in the conversion procedure.
To verify the feasibility of \textit{e}Sampling ADCs, we present a circuit-level design using standard complementary metal oxide semiconductor (CMOS) 65 nm technology. An  \textit{e}Sampling 8-bit ADC which samples at 40 MHZ  is designed on a Cadence Virtuoso platform. Our experimental study involving Nyquist rate sampling of bandlimited signals demonstrates that such ADCs are indeed capable of harvesting more energy than that spent during analog-to-digital conversion, without affecting the accuracy.

\end{abstract}

\vspace{-0.2cm}
\section{Introduction}

Physical signals are analog in nature, taking values in continuous sets over a continuous time interval.
In order to process and extract information from such signals using digital hardware, they must be accurately represented in digital form. 
Analog-to-digital converters (ADCs) thus play an important role in digital signal processing systems \cite{eldar2015sampling}. 
ADCs are typically a major source of energy consumption, as their power dissipation grows with the sampling rate and the quantization resolution, and thus their ability to accurately represent the acquired signal is typically limited by the available power \cite{4403893}. Nowadays, ADCs are utilized in a multitude of energy-limited systems, including communication devices \cite{shi2002data}, wireless sensors \cite{jain2018ideg}, and medically implanted devices \cite{6774474}.  Therefore, there is a growing need for ADCs capable of reliably acquiring signals while consuming low power. 



The existing strategies proposed in the literature to facilitate energy efficient acquisition of analog signal can be divided into those taking a signal processing approach, and techniques focusing on circuit level design. Signal processing approaches typically aim for allowing the ADC to operate at reduced sampling rate and quantization resolution by accounting for how the acquired signal is processed and prior information on the signal itself
\cite{michaeli2011xampling, cohen2018analog, cohen2018sub, shlezinger2019joint, jain2018ideg}.
Additionally, in scenarios where the signal is acquired for some task, i.e., to recover some underlying information, it was recently shown that the desired information could be accurately recovered from the output of low-resolution ADCs by properly designing the acquisition system \cite{shlezinger2018hardware, shlezinger2019deep, shlezinger2020learning, shlezinger2020task}. 
An alternative signal processing oriented method which does not limit the rate and resolution of ADC is based on acquiring a portion of the analog signal to be processed while utilizing the remaining part for energy harvesting. This strategy, typically studied in the context of communication receivers as simultaneous wireless information and power transfer (SWIPT), considers time or power splitting of the analog signal \cite{6489506,6503739,liu2013wireless,lu2014wireless}. However, it induces some inevitable loss on the system performance as only a portion of the signal is converted into a digital representation.  
These aforementioned signal processing methods typically focus on the signal model and the task for which it is acquired, without accounting for the ADC circuitry.


Circuit level methods rely on the hardware architecture of ADC devices. The circuit level approach generally considers designing energy efficient ADC circuitry, which is capable of operating with reduced power consumption. This can be achieved by reducing the circuit power supply \cite{6774474} and/or limiting the operating frequency \cite{artan2012optimizing} in order to reduce the overall power consumption.  
An alternative technique is to modify the circuit components in existing ADC architectures and combine various designs in the acquisition, such as sample-and-hold (S/H) ADCs, flash ADCs, sigma-delta ADCs, and time-interleaved ADCs, to improve their energy efficiency, see, e.g.,  \cite{6936944,8727467,8676062, 8017456}. Such circuit-oriented designs which focus on the hardware aspects of acquisition, do not account for the model of the analog signal and the task for which it is acquired.

A popular power efficient ADC is the S/H based successive approximation register (SAR) architecture, which is capable of operating at high resolution and a small form factor with relatively low power consumption \cite{razavi1995principles}. The power consumption of SAR ADCs can be further reduced by incorporating energy efficient switching schemes, as proposed in \cite{hariprasath2010merged, liu201010}. 
In S/H architectures, the circuit used to sample the input analog signal consists of two phases, \textit{acquisition phase} and \textit{hold phase} in each sampling period. In the acquisition phase, the S/H circuit tracks the input analog signal. The sampled value captured in the acquisition phase is then converted into digital form, i.e., a sequence of bits, during hold phase. Therefore, during the sampling process of S/H ADCs, the input signal is processed only for a fraction of the overall sampling period (acquisition phase) and is neglected/discarded for the remaining time interval (hold phase) \cite{1464555,mccreary1975all}.
The fact that the signal is not accessed in a dominant portion of the sampling period, motivates the extension of S/H ADCs, and particularly S/H SAR ADCs, to continuously utilize the analog signal in order to mitigate power consumption. 

In this work, we combine signal processing tools with circuit level methods to propose an {\em \textit{e}Sampling ADC}, which harvests energy from the acquired signal while converting it into a digital representation. The \textit{e}Sampling ADC builds upon the S/H ADC architecture while introducing an additional energy harvesting circuit. In the resulting architecture, the signal is harvested during hold phase, i.e., when it is not utilized in conventional S/H ADCs.  This operation allows \textit{e}Sampling ADCs to harvest energy from the sampled signal without altering the conversion procedure. Our analysis of \textit{e}Sampling ADCs  formulates the theoretical foundations for joint acquisition and energy harvesting, and generalizes the experimental results of our previous work  \cite{jainExp2020}, which demonstrated that energy harvesting can be combined with sensing circuits. As opposed to SWIPT systems, in which the overall operation of the system is modified to allow energy harvesting while conventional ideal ADCs are assumed \cite{lu2014wireless}, \textit{e}Sampling  exploits an inherent property of ADC devices to harvest energy as a natural byproduct of their hardware architecture.  This makes \textit{e}Sampling an attractive technology which can be easily incorporated into existing devices.

Our theoretical study of \textit{e}Sampling ADCs analyzes its potential in terms of the ability to harvest energy while maintaining a desired accuracy of signal reconstruction.
To that aim, we focus on the acquisition of stationary random processes and characterize the resulting tradeoff between the ability to accurately reconstruct the signal from its samples and the energy harvested from it, referred to henceforth as the {\em energy-fidelity tradeoff}. Our analysis identifies how to set the sampling rate to optimize this tradeoff when operating under energy constraints or fidelity restrictions on the reconstruction.
The results allow us to numerically characterize the maximal accuracy in which  any signal can be \textit{e}Sampled using only harvested energy, i.e., without requiring any energy from its power source.  The energy consumed in acquisition is  determined by the specific components comprising the ADC circuit. We show that \textit{e}Sampling ADCs operating with a typical set of ADC parameters are capable of fully reconstructing signals of various power spectral density (PSD) profiles with negligible distortion, while harvesting at least as much energy as they consume.  In particular, we show that an \textit{e}Sampling ADC with $12$ bits quantization can acquire a bandlimited signal at the Nyquist rate while harvesting more energy than it consumes. 

We then proceed to illustrate the hardware feasibility of such a device. To that aim, we design the circuitry of an \textit{e}Sampling 8-bit SAR ADC which samples at 40~MHz on 65~nm complementary metal oxide semiconductor (CMOS) technology, and provide guidelines for setting its parameters to achieve a desired amount of harvested energy. The experimental evaluation of the \textit{e}Sampling SAR ADC circuit, carried out on the Cadence Virtuoso platform, shows that the amount of energy harvested can be much larger than the amount of energy consumed during the conversion procedure. This is achieved without affecting the signal reconstruction accuracy when acquiring a bandlimited signal while satisfying Nyquist condition. Our experiment indicates that the theoretical potential of \textit{e}Sampling can be translated into an actual ADC circuit, which accurately acquires analog signals while harvesting more power than it consumes.

The rest of this paper is organized as follows: In Section~\ref{sec:Model}, we present our \textit{e}Sampling system model. Section~\ref{sec:Analysis} analyzes the associated energy-fidelity tradeoff. The circuit-level design and its experimental study are presented in Section~\ref{sec:circuit_level_implementation}. Finally, Section~\ref{sec:Conclusions} provides concluding remarks.

\begin{figure*}[t]
\centering
\includegraphics[width=\linewidth]{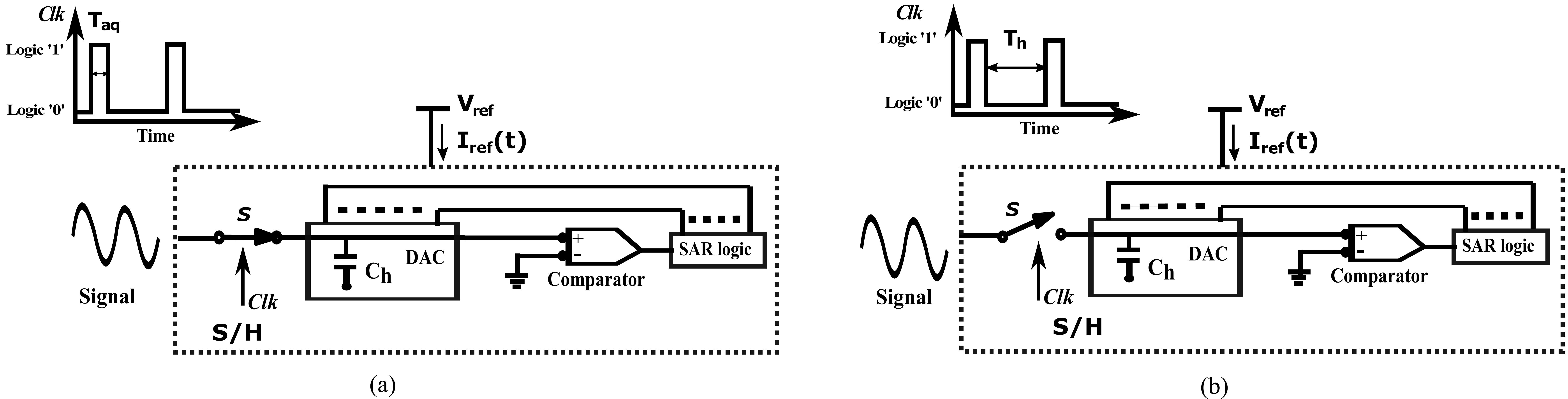}
\caption{S/H SAR ADC illustration: (a) acquisition phase (b) hold phase.}
\label{fig:1}
\end{figure*}

\section{System Model}
\label{sec:Model}
In this section, we detail the proposed ADC model from a high-level perspective. We begin by briefly reviewing S/H-based SAR ADCs and their associated energy consumption in Subsection \ref{subsec:SH}. Then, we present how S/H ADCs can be extended into \textit{e}Sampling ADCs which harvest energy in addition to signal acquisition in Subsection \ref{subsec:eSampling}. 

\subsection{Sample-and-Hold ADC Model}
\label{subsec:SH}
\subsubsection{High-level description}
S/H is a common ADC architecture. Such ADCs acquire each sample in two phases, determined by a switch $S$, as illustrated in Fig. \ref{fig:1}: In the acquisition phase, the signal is connected to a capacitor $C_{\rm h}$, referred to as a holding capacitor,  which is charged to the input analog voltage, as depicted in Fig. \ref{fig:1}(a). The time required by the holding capacitor to charge to the input voltage, which dictates the acquisition time, is given by  \cite{razavi1995principles}
\begin{align}
   T_{\rm aq}& =\alpha_\tau R_{\rm on} C_{\rm h}, \label{eq:Taq}
\end{align} 
where $R_{\rm on}$ is the on-resistance of the switch $S$, and $\alpha_\tau$ is the number of time constants, i.e.,  $R_{\rm on} C_{\rm h}$ required for the capacitor to be fully charged.

Once the acquisition phase is over, the hold phase begins, in which the discrete sample, i.e., the voltage stored in the holding capacitor, is quantized into digital bits. 
During hold phase, whose duration is denoted by $T_{\rm h}$, the input signal is disconnected from the S/H circuit and $C_{\rm h}$ holds the acquired voltage to accomplish the successful conversion of the acquired sample into digital bits as illustrated in Fig. \ref{fig:1}(b). Both $T_{\rm h}$ and $C_{\rm h}$, must be set to allow the quantization circuit of the ADC to complete the conversion. 

When the  quantizer is based on SAR logic, the overall architecture is referred to as a SAR ADC. An $n$-bit SAR ADC consists of a comparator, digital-to-analog converter (DAC), and a SAR logical circuit which successively refines the digital representation. 
To allow successful quantization into $n$ bits,
the hold time required to quantize each sample must satisfy  \cite{mccreary1975all}
\begin{equation}
  T_{\rm h}\geq n \alpha_\tau R_{\rm q} C_{\rm h},  \label{eq:Th}
\end{equation}
where $R_{\rm q}$ is the equivalent resistance of the quantizer binary scale switches. 
Therefore, the sampling period, i.e., the duration of acquiring a single sample, is lower bounded by the following expression
\begin{equation}
    T_{\rm s} = T_{\rm aq}+T_{\rm h}\geq (R_{\rm on}+nR_{\rm q})\alpha_\tau C_{\rm h}. \label{eq:Ts}
\end{equation}

In S/H SAR ADCs, the on-resistance of the switch $R_{\rm on}$ is commonly not larger than the resistance of the quantizer binary scale switches $R_{\rm q}$. Thus from \eqref{eq:Taq} and \eqref{eq:Th}, it is evident that $T_{\rm h}$ is typically much larger than $T_{\rm aq}$, particularly when using high resolution quantizers, such as ADCs with $n \geq 8$ bits. Consequently, the input signal, which is tracked only during the acquisition phase, is discarded during most of the sampling period. 

\subsubsection{Energy consumption}
In general, the energy consumption of a circuit is typically a function of the time duration it is active, and the amount of power drawn from the supply, denoted here by $V_{\rm ref}$. As $T_{\rm h}$ is typically much larger than $T_{\rm aq}$, most of the energy required by S/H SAR ADCs is consumed during hold phase \cite{mccreary1975all, hariprasath2010merged}. 

In particular, the only energy consumed during acquisition phase, denoted $E_{\rm aq}$,  is  that needed to toggle the sampling switch $S$. In contrast, the energy consumption during hold phase, denoted $ E_{\rm hold}$, is comprised of the energy used by each of the components taking part in the quantization:
\begin{equation}
    E_{\rm hold}=E_{\rm DAC}+E_{\rm c}+E_{\rm sl}, \label{53}
\end{equation}
where $E_{\rm DAC}$, $E_{\rm c}$, and $E_{\rm sl}$ are the energy consumption of the DAC array, comparator, and SAR logic, respectively. Consequently, $E_{\rm hold}$ effectively represents the power consumed per sample by S/H SAR ADCs \cite{mccreary1975all, hariprasath2010merged}. We elaborate on the quantities in \eqref{53}, which are dictated by the specific circuit parameters used, in Section \ref{sec:circuit_level_implementation} where a concrete circuit-level design is discussed. Here, we note that $E_{\rm hold}$ typically takes the form of a second-order polynomial in the reference voltage $V_{\rm ref}$ \cite{6043594}, i.e., 
\begin{equation}
\label{eqn:EholdExp}
    E_{\rm hold}= a_1(n)V_{\rm ref} + a_2(n)V_{\rm ref}^2.
\end{equation}
The coefficients $a_1(n)$ and $a_2(n)$ in \eqref{eqn:EholdExp} are positive constants determined by the number of bits $n$ and the quantization circuit parameters, and can grow dramatically with  $n$. This makes energy consumption a major bottleneck of high resolution ADCs, motivating the proposed \textit{e}Sampling architecture detailed next.

\subsection{\textit{e}Sampling ADC Architecture} 
\label{subsec:eSampling}
As mentioned above, during hold phase, the capacitor $C_{\rm h}$ holds the acquired voltage sample, which is converted into a set of digital bits. In this interval, the input signal is disconnected from the circuit by the switch $S$.  
In order to mitigate the energy consumption of S/H SAR ADCs without modifying their sampling and quantization procedure, we propose to harvest the input signal energy  by connecting it to an energy harvesting circuit during the hold phase, as illustrated in Fig.~\ref{fig:eSampling_system_model}. 
Henceforth, the proposed architecture is referred to \textit{e}Sampling ADC. 
 
As depicted in Fig.~\ref{fig:eSampling_system_model}, the energy harvesting capability is enabled by passing the signal observed during hold time through a conditioning circuit, whose output is used to charge an energy harvesting capacitor $C_{\rm EH}$ to a voltage level $V_{\rm EH}$. 
The energy harvesting circuit can be designed using passive elements, as we do in our proposed design detailed in Section~\ref{sec:circuit_level_implementation}. Hence, no external power supply is required \cite{7404333}. 
The purpose of the signal conditioning circuit used in energy harvesting devices is to facilitate the storage of the energy of the signal in the capacitor $C_{\rm EH}$ \cite{8410909,8740972, 7081797}. For instance, a rectifier can act as a signal conditioning circuit, reducing fluctuations in the amount of energy harvested in the presence of alternating signals.  Similarly, voltage regulator circuits and DC-DC step up converters can also be used to enhance the overall efficiency of the energy harvesting system \cite{sarker2013designing}. 
The common measure for the quality of an energy harvesting circuit is the efficiency parameter, denoted by $\eta \in [0,1]$, which represents the fraction of the energy of the input signal that is harvested. Finally, in order to connect the input signal to the quantization circuit during acquisition time and to the energy harvesting circuit during hold time, the sampling switch $S$ is replaced by a two-way switch $\tilde{S}$. A possible circuit design such a two-way switch is detailed in Section \ref{sec:circuit_level_implementation}.


The amount of energy consumed in acquisition phase given in \eqref{eqn:EholdExp} is dictated by the design parameters of the circuitry, which also affect the sampling rate via \eqref{eq:Ts}. In particular, the sampling duration is the sum of the acquisition time $T_{\rm aq}$ and the hold time $T_{\rm h}$. Further,  the amount of time during which energy is harvested from the input signal per sampling period is at most $T_{\rm h}$. Recalling that typically $T_{\rm h} \gg T_{\rm aq}$, a significant portion of the sampling interval can be allocated for harvesting energy from the input signal. Since energy is only harvested during hold time, in which conventional S/H ADCs do not utilize the analog signal, the ability to harvest energy in \textit{e}Sampling ADCs {\em does not affect the acquisition operation}. Specifically, for a given sampling rate, \textit{e}Sampling ADCs implement the same conversion mapping as standard S/H ADCs operating at the same rate. Nonetheless, \textit{e}Sampling provides the ability to trade acquisition accuracy for harvesting more energy.  This is due to the fact that increasing the sampling interval allows \textit{e}Sampling ADCs to dedicate more time to energy harvesting, possibly at the cost of degrading the accuracy in reconstructing the analog signal from its digital representation. 

 The goal of our analysis of \textit{e}Sampling ADCs presented in the following section is to quantify the theoretical potential benefits of such an architecture, which is capable of simultaneously acquiring analog signals into a digital form while harvesting their energy. Both the amount of energy consumed in conversion and that harvested in \textit{e}Sampling are determined by the specific circuitry, encapsulated in \eqref{eqn:EholdExp} and the energy efficiency parameter $\eta$, respectively. Therefore, in our analysis we fix the circuit parameters, e.g., $a_1(n)$, $a_2(n)$, $\eta$, etc., and express how the accuracy in reconstruction and the amount of energy harvested vary as the sampling interval changes. We are particularly interested in characterizing the amount of energy harvested in the regime in which the distortion induced by S/H conversion is negligible, e.g., Nyquist rate sampling of bandlimited signals, and understanding when is it possible for  \textit{e}Sampling ADCs  to operate at this regime while harvesting at least as much energy as they consume. Our theoretical analysis detailed in the sequel reveals that such a regime of operation is indeed feasible with typical ADC circuit parameters when acquiring bandlimited signals  while using up to $12$ bits per sample.

\section{\textit{e}Sampling ADC Analysis}
\label{sec:Analysis}
 In this section, we analyze the capabilities of the proposed \textit{e}Sampling ADC in terms of  the amount of energy one can harvest while meeting a given level of reconstruction accuracy, as well as the achievable accuracy for harvesting a desired amount of energy. The interplay between these key performance measures is determined by the selection of the sampling rate, as we show in the following.
We begin by formulating the signal model under which our analysis is carried out, and the corresponding problem of characterizing the associated energy-fidelity tradeoff, which arises from the \textit{e}Sampling ADC paradigm in Subsection \ref{subsec:Problem}. Then, we derive the achieved normalized mean-squared error (NMSE) under the considered model in Subsection \ref{subsec:NMSE}. The derived NMSE is used to characterize the energy-fidelity tradeoff in Subsection \ref{subsec:tradeoff}, and to obtain as a special case the maximal amount of energy which can be harvested when sampling a bandlimited signal at a rate satisfying the Nyquist condition, i.e., allowing perfect recovery.  We demonstrate a few examples of energy-fidelity tradeoff curves for signals with different spectral profiles in Subsection~\ref{subsec:Examples}. Finally, we discuss the pros and cons of \textit{e}Sampling ADC in light of our analysis in Subsection~\ref{subsec:discussion}. 


\subsection{Problem Formulation}
\label{subsec:Problem} 
The \textit{e}Sampling ADC detailed in Subsection \ref{subsec:eSampling} harvests energy during hold phase. This implies that more energy can be harvested by increasing the hold time, which in turn increases the sampling period, potentially degrading the ability to reconstruct the signal from its samples. Therefore, to unveil the potential of \textit{e}Sampling ADCs, we first wish to analyze the fundamental tradeoff between the amount of energy harvested in \textit{e}Sampling and the resulting fidelity in signal reconstruction. We are particularly interested in: $1)$ Quantifying the maximum amount of energy that could be harvested when acquiring bandlimited signals at the Nyquist rate, i.e., without compromising the signal reconstruction accuracy; and $2)$ Characterizing the achievable NMSE when the ADC harvests at least as much power as it consumes. 

\begin{figure}
\centering
\includegraphics[width=\figWidth]{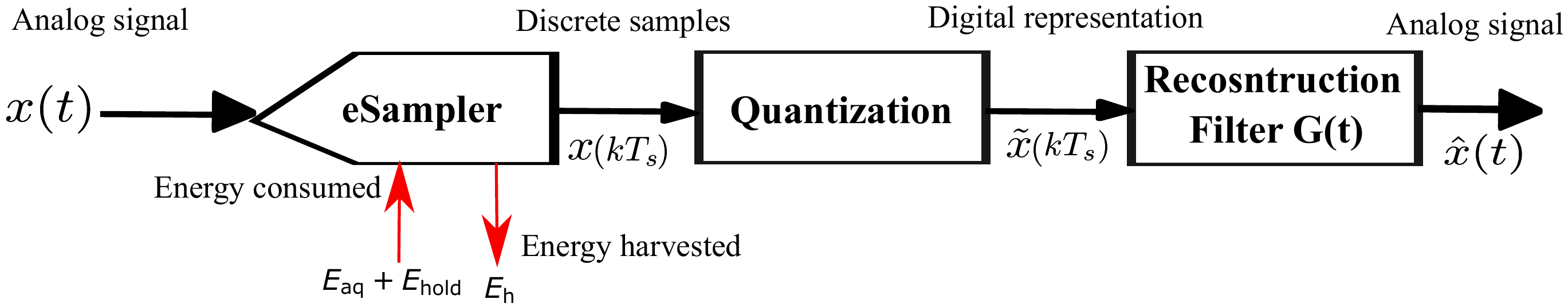}
\caption{Acquisition and reconstruction via \textit{e}Sampling ADC illustration.}
\label{fig:2}
\end{figure}

In the analysis carried out in this section we consider a stochastic input signal $x(t)$ modeled as a zero-mean wide sense stationary (WSS) process, with  variance $\sigma^2_x$, and  PSD $S_x(f)$. The signal $x(t)$ is sampled uniformly with sampling interval $T_{\rm s}$, resulting in the discrete-time signal $x(k T_{\rm s})$, $k \in \mathcal{Z}$, where  $\mathcal{Z}$ is the set of integers. The sampled series is quantized with $n$ bits per sample into the digital sequence $\Tilde{x}(k T_{\rm s})$. The digital representation is utilized to recover the analog signal $x(t)$ using a linear reconstruction filter $G(t)$,  which is designed to minimize the NMSE between $x(t)$ and the recovered signal $\hat{x}(t)$ as in \cite{michaeli2008high,shlezinger2019joint}. The reconstructed signal is
\begin{equation}
\hat{x}(t) = \sum_{k \in \mathcal{Z}} G(t-kT_{\rm s}) \tilde{x}(kT_{\rm s}).  \label{10}
\end{equation}
The overall system is illustrated in Fig. \ref{fig:2}.
 
\begin{figure}
\centering
\includegraphics[width=\figWidth]{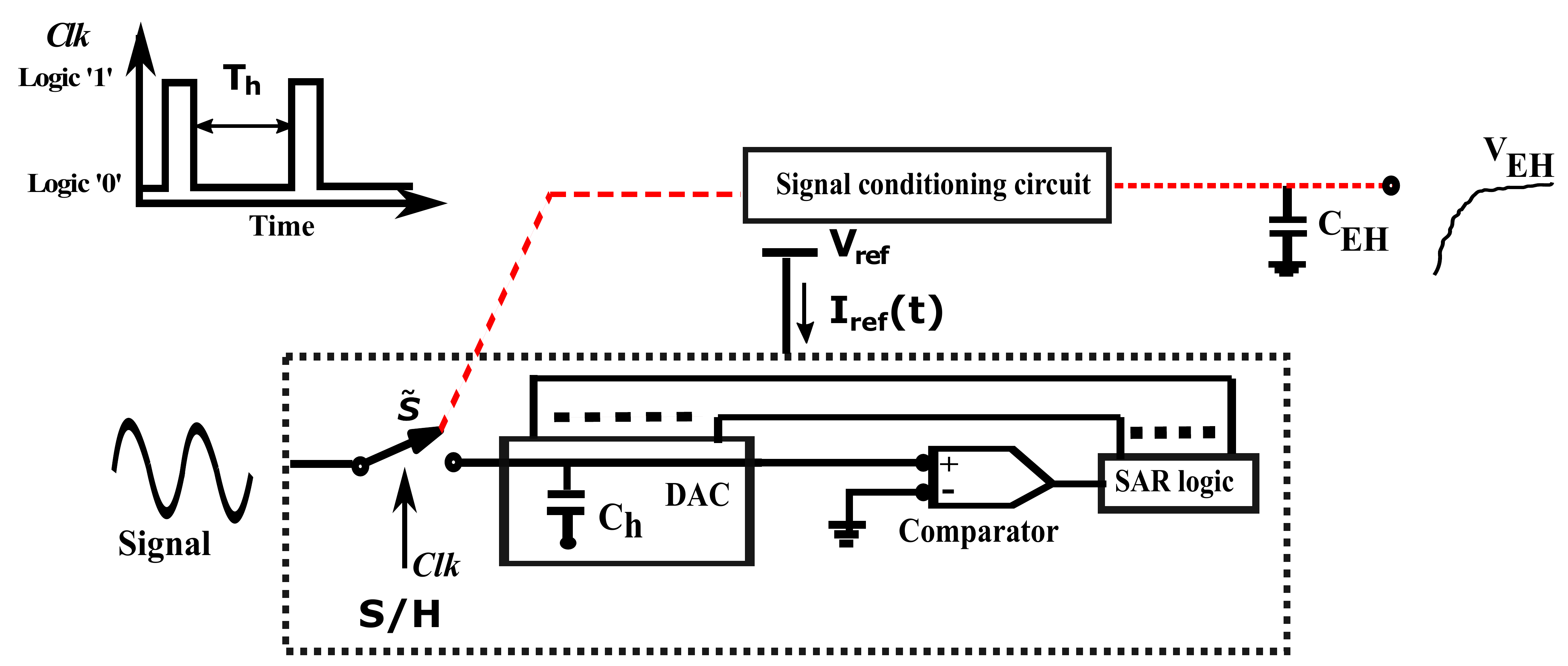}
\caption{Proposed \textit{e}Sampling ADC system model.}
\label{fig:eSampling_system_model}
\end{figure}

The NMSE in reconstructing $x(t)$ from $\hat{x}(t)$ 
is given by 
 \ifFullVersion
\begin{align}
    \zeta =\frac{1}{\sigma_x^2 T_{\rm s}}\int_{0}^{T_{\rm s}}\E\{|x(t)-\hat{x}(t)|^2\}dt, \label{11}
    \end{align}
\else
$\zeta =\frac{1}{\sigma_x^2 T_{\rm s}}\int_{0}^{T_{\rm s}}\E\{|x(t)-\hat{x}(t)|^2\}dt$, 
 \fi 
where $\E\{\cdot\}$ is the stochastic expectation.
The amount of expected energy harvested per sampling period is given by
\begin{align}
    E_{\rm h}=\eta \frac{1}{R_{\rm h}} \int_{T_{\rm aq}}^{T_{\rm s}}\E\{|x(t)|^2\} dt = \frac{\eta}{R_{\rm h}} T_{\rm h}\sigma_x^2,  \label{expected_energy_harvested}
\end{align}
where $\eta$ and $R_{\rm h}$ is the efficiency and the resistance of the energy harvesting circuit, respectively. As mentioned above, the energy harvesting circuit is comprised of passive elements, and does not require an external power source. Therefore, the overall energy consumption per sample using the proposed \textit{e}Sampling ADC can be given as $E_{\rm aq}+E_{\rm hold}-E_{\rm h}$ as illustrated in Fig. \ref{fig:2}. Recall that the overall energy consumption is typically dominated by the energy used during hold phase, i.e.,  $E_{\rm aq} \ll E_{ \rm hold}$, and hence the ratio of the amount of energy harvested to the energy consumption per sample can be approximated as
    $E_{\rm ratio}= \frac{E_{\rm h}}{E_{\rm hold}}$. 
The value of $E_{\rm hold}$ is dictated by the power supply voltage $V_{\rm ref}$ and the number of quantization bits $n$, as well as the SAR architecture and circuit parameters, as we show for our design detailed in Section \ref{sec:circuit_level_implementation}.  

In the following subsections, we study the fundamental tradeoff between the reconstruction accuracy, modelled as the NMSE, and the portion of the energy consumed in analog-to-digital conversion to that harvested by \textit{e}Sampling, referred to as the {\em energy-fidelity tradeoff}. To trade energy efficiency for fidelity, we modify the sampling rate for a fixed quantization resolution $n$ and fixed acquisition time $T_{\rm aq}$. The reconstruction accuracy can be improved by increasing the sampling rate, however \textit{e}Sampling ADC will harvest less energy, and hence the inherent tradeoff between these parameters. In particular, we focus on ADCs operating with relatively high resolution, where energy consumption constitutes a major challenge. 
The following analysis sheds light on the potential of joint acquisition and energy harvesting. For example, it quantifies the minimal recovery NMSE which allows a fixed $n$-bit ADC to operate at zero power, i.e., $E_{\rm ratio}=0~\text{dB}$. Alternatively, it allows identifying the quantization resolution $n$ for which the \textit{e}Sampling ADC can sample a bandlimited signal at Nyquist condition and operate at zero power. For instance, we use our results to show that bandlimited signals can be \textit{e}Sampled at Nyquist rate with up to $12$ bits per sample while harvesting more energy than that consumed. Finally, the characterization of the energy-fidelity tradeoff allows computing the maximal amount of energy which can be harvested for an allowed level of reconstruction accuracy for both bandlimited and non-bandlimited signals, as a function of the ADC circuitry parameters.

%
%
 
\subsection{Reconstruction NMSE}
\label{subsec:NMSE}
In general, the NMSE depends on both the sampling rate as well as the quantization resolution \cite{kipnis2018fundamental}. Since we focus on relatively high rate quantization, the NMSE due to quantization is well-approximated by the $6$ dB rule-of-thumb \cite[Ch. 23]{polyanskiy2014lecture}, and is thus on the order of  $10^{-0.6n}$ \cite{razavi1995principles},  
resulting in a negligible contribution to the overall NMSE of less than roughly $10^{-5}$ for $n \geq 8$.  
Therefore, henceforth the focus is on the the NMSE between $x(t)$ and $\hat{x}(t)$ due to the sampling procedure alone, expressed in the following theorem, derived in  \cite{michaeli2008high}: 
 
\begin{theorem}
\label{thm:MSE}
The minimal achievable NMSE in reconstructing a uniformly sampled WSS signal $x(t)$ with sampling frequency $f_s = 1/T_{\rm s}$ using a linear reconstruction filter, $G(t)$ is 
\begin{align}
\zeta(T_{\rm s}) =& 1-\frac{1}{\sigma_x^2} \sum_{k\in \mathcal{Z}} \int_{-\frac{f_s}{2}}^{\frac{f_s}{2}}\frac{|S_{x}(f-kf_s)|^2}{\sum_{k'\in \mathcal{Z}} S_{x}(f-k'f_s)}df. \label{57}
\end{align}
\end{theorem}

\smallskip
 To achieve \eqref{57}, the linear recovery filter $G(t)$ in \eqref{10} is set according to \cite{michaeli2008high,shlezinger2019joint}, i.e., its frequency response $\mathcal{F}(G)(f)$ should be set to $\mathcal{F}(G)(f) = \frac{S_x(f)}{\sum_{k \in \mathcal{Z}}S_x(f-k f_s)}$, where $\mathcal{F}(\cdot)$ denotes the Fourier transform. This digital filter setting  results in the minimal achievable NMSE between $x(t)$ and $\hat{x}(t)$.
Theorem \ref{thm:MSE} generalizes the celebrated Shannon-Nyquist theorem, 
as stated in the following corollary:
\begin{corollary}
\label{cor:NyqSamp}
When $x(t)$ is bandlimited and the sampling frequency satisfies Nyquist condition, the resulting NMSE is zero.
\end{corollary}
\begin{IEEEproof}
If $x(t)$ is bandlimited, then there exists some finite $f_m$ such that $S_x(f) =0$ for all $|f| > f_m$. 
When the sampling rate satisfies Nyquist condition, then $f_s \geq 2 f_m$. Consequently, the summands in \eqref{57} are non-zero only at $k=k'=0$, and hence
\begin{align}
    \zeta(1/f_s) &=1-\frac{1}{\sigma_x^2} \int_{-\frac{f_s}{2}}^{\frac{f_s}{2}}\frac{|S_{x}(f)|^2}{S_{x}(f)}df \notag \\
    &= 1-\frac{1}{\sigma_x^2} \int_{-f_m}^{f_m}\frac{|S_{x}(f)|^2}{S_{x}(f)}df  =0,
\end{align}
proving the corollary. 
\end{IEEEproof}

\smallskip
We next give an example of how Theorem~\ref{thm:MSE} is computed:
\begin{example}
\label{exm:Flat}
Consider a bandlimited signal whose spectral support is $[-f_{\rm m},f_{\rm m}]$ for some $f_{\rm m}>0$ with flat PSD. The obtained NMSE for such signals computed via Theorem \ref{thm:MSE} is given by
\begin{equation}
  \zeta(1/f_s) = \begin{cases}
  1-\frac{f_s}{2f_{\rm m}} & f_{\rm s} \le 2f_{\rm m},\\ 
0 & {\rm otherwise}. \label{eq:nmse_flat_psd}
  \end{cases}
\end{equation}
\end{example}

\begin{figure*}
\centering
\includegraphics[width=0.8\linewidth]{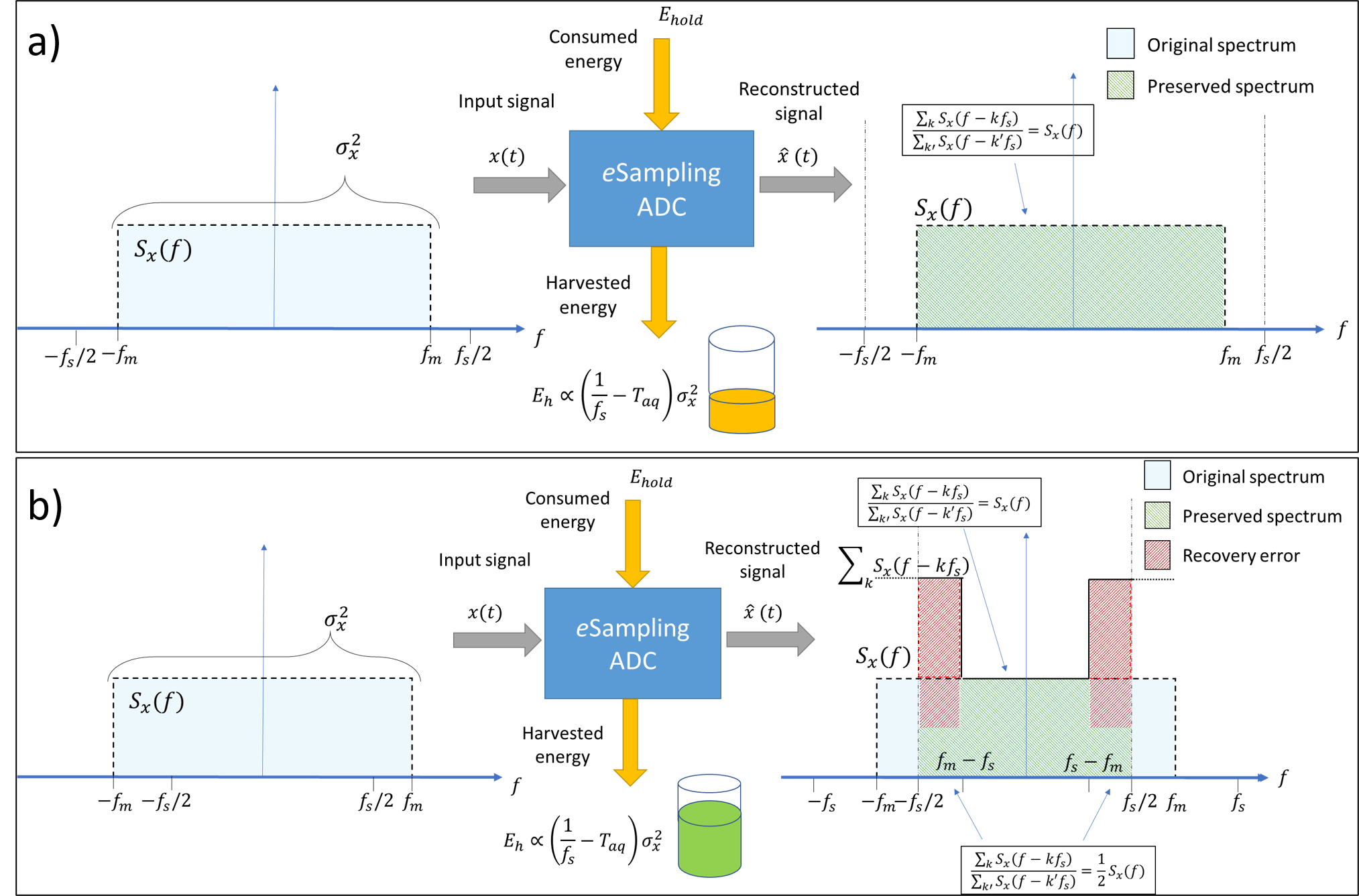}
\caption{Illustration of \textit{e}Sampling of a signal with a flat PSD for: (a) Sampling at Nyquist rate, while harvesting an amount of energy proportional to $T_{\rm h} = 1/f_s - T_{\rm aq}$; (b) Sampling at sub-Nyquist rate, thus trading recovery accuracy for harvesting more energy.}
\label{fig:drawing}
\end{figure*}

Fig. \ref{fig:drawing} illustrates of the recovery NMSE result in Theorem~\ref{thm:MSE}, showing which spectral portions of a signal with a flat PSD as in Example \ref{exm:Flat} are preserved by the NMSE minimizing reconstruction. In particular, Fig. \ref{fig:drawing} demonstrates how the complete spectrum is preserved when sampling above Nyquist rate, while sub-Nyquist sampling yields some recovery error due to aliased components. Fig. \ref{fig:drawing} also depicts the amount of energy harvested from the signal based on \eqref{expected_energy_harvested}, showing that reduction in the sampling rate allows to harvest more energy in \textit{e}Sampling at the cost of less accurate recovery, leading to the energy-fidelity tradeoff of \textit{e}Sampling analyzed in the sequel. 

\subsection{Energy-Fidelity Tradeoff}
\label{subsec:tradeoff}
In order to express the energy consumed in acquisition, we must first specify the voltage of the power supply $V_{\rm ref}$. This value should be larger than the amplitude of the input signal with high probability to avoid overloading the ADC. Consequently, in the following we write the value of $V_{\rm ref}$ as some multiple $K>1$ of the input standard deviation, i.e., the supply voltage is written as $V_{\rm ref} = K \sigma_x$. This general formulation allows us to relate the reference voltage with the overload probability of the quantizer, since the overload probability satisfies $P(|x(t)|\geq V_{\rm ref})\leq K^{-2}$ by Chebyshev's inequality \cite{shlezinger2018hardware}. Therefore, the ratio between the expected energy harvested \eqref{expected_energy_harvested} and consumed \eqref{eqn:EholdExp} for \textit{e}Sampling of a WSS signal can be written as
\begin{align}
    \!\!E_{\rm ratio} \!= \!    \frac{\frac{\eta}{R_{\rm h}} (T_{\rm s}-T_{\rm aq})\sigma_x^2}{a_2(n)K^2\sigma_x^2 \!+\! a_1(n) K \sigma_x}.\label{eq:eratio_WSS}
\end{align}

 Recall that for a fixed sampling interval, \textit{e}Sampling ADCs implement the same conversion mapping as conventional S/H ADCs. Consequently, when one does not account for the distortion induced in quantization as we do here, WSS signals acquired by an \textit{e}Sampling ADC operating with sampling interval $T_s$ can be recovered with the NMSE $\zeta(T_s)$ stated in Theorem \ref{thm:MSE}.
We therefore use the  expressions for the achievable NMSE \eqref{57} and the energy ratio \eqref{eq:eratio_WSS} to characterize the energy-fidelity tradeoff of \textit{e}Sampling.  

Under the considered setting, we formulate how the recovery accuracy and the energy ratio behave as the sampling period $T_{\rm s}$ varies. Recalling that the acquisition time $T_{\rm aq}$ is determined by the ADC circuit parameters \eqref{eq:Taq}, modifying the sampling period is equivalent to tuning the hold time $T_{\rm h}$. The energy-fidelity  tradeoff of \textit{e}Sampling is thus encapsulated in two complementary optimization problems: 
The first aims at finding the minimal achievable NMSE under a given energy constraint $\delta >0$, i.e., 
%
\ifFullVersion
\begin{align}
  &\zeta\opt (\delta)  =\mathop{\min}\limits_{T_{\rm s} > T_{\rm aq}} {\zeta}, \label{58} \\
    &\text{subject to } E_{\rm ratio} \geq \delta. \notag 
\end{align}
\else
\begin{equation}
   \zeta\opt (\delta)  =\mathop{\min}\limits_{T_{\rm s} > T_{\rm aq}} {\zeta} \ \ \ \ \ \ \ \text{   s.t.}\ E_{\rm ratio} \geq \delta. \label{58}
\end{equation}
\fi
%
%
Setting $\delta = 0$ dB, implies that  $E_{\rm hold}=E_{\rm h}$. Therefore, solving \eqref{58} with  $\delta = 0$ dB reveals the minimal NMSE achievable by an \textit{e}Sampling ADC which harvests at least as much energy as it consumes, i.e., when operating at zero power. A positive value of $\delta$ (in dB) implies an energy saving ADC which harvests more energy than its consumption per sample, namely, converting the signal only adds power to the system. 

An alternative formulation seeks to maximize the energy harvested under a given fidelity constraint $\epsilon > 0$, i.e., 
\ifFullVersion
\begin{align}
   &E_{\rm ratio}\opt(\epsilon) = \mathop{\max}\limits_{T_{\rm s} > T_{\rm aq}} E_{\rm ratio}, \label{59} \\
    &\text{subject to } {\zeta} \le \epsilon. \notag 
\end{align}
\else
\begin{equation}
   E_{\rm ratio}\opt(\epsilon) = \mathop{\max}\limits_{T_{\rm s} > T_{\rm aq}} E_{\rm ratio}  \ \ \ \ \ \text{   s.t.}\ {\zeta} \le \epsilon. \label{59}
\end{equation}
\fi
 For instance, consider a bandlimited signal. In such a case, one can achieve $\zeta = 0$ by \textit{e}Sampling at Nyquist rate, and harvest energy ratio $E_{\rm ratio}^{\rm opt}(0)$, i.e., the maximal ratio of the harvested to energy to the consumed one when seeking ideal recovery. For non-bandlimited signals, approaching zero NMSE generally requires infinitesimally small sampling interval, which is not feasible due to the lower bound on $T_s$ dictated by the ADC circuity in \eqref{eq:Ts}. Consequently, when acquiring non-bandlimited signals (or extremely wideband signals), one would typically be more interested in evaluating \eqref{59} for some small yet feasible NMSE bound $\epsilon>0$.  


Problems \eqref{58}-\eqref{59} allow to characterize the energy-fidelity tradeoff, stated in the following theorem:
\begin{theorem}
\label{thm:Tradeoffs}
\begin{subequations}
\label{eqn:Tradeoffs}
Let $T_{\rm h}(\delta)$ be given by 
\begin{equation*}
    T_{\rm h}(\delta) :=\frac{\delta R_{\rm h}}{\eta \sigma_x^2} \left(a_2(n) K^2\sigma_x^2 \!+\! a_1(n) K \sigma_x \right).
\end{equation*}
By setting $f_{s}(\delta) = \frac{1}{T_{aq}+T_{\rm h}(\delta)}$, the solution to \eqref{58} is  
\begin{equation}
    \label{eqn:Tradeoff1}
    \zeta\opt(\delta)\! =\! 1\! -\!\frac{1}{\sigma_x^2} \sum_{k\in \mathcal{Z}} \int_{-\frac{f_{s}(\delta)}{2}}^{\frac{f_{s}(\delta)}{2}}\frac{|S_{x}(f\!-\!kf_{s}(\delta))|^2}{\sum_{k'\in \mathcal{Z}} S^H_{x}(f\!-\!k'f_{s}(\delta))}df.
\end{equation}
Similarly, by letting $T_{\rm s}(\epsilon)$ be the maximal sampling interval satisfying $\zeta(T_{\rm s}(\epsilon)) = \epsilon$ in \eqref{57}, then the solution to \eqref{59} is
\begin{align}
    \!\!E_{\rm ratio}\opt\!(\epsilon) \!= \!    \frac{\frac{\eta}{R_{\rm h}} (T_{\rm s}(\epsilon)-T_{\rm aq})\sigma_x^2}{a_2(n) K^2\sigma_x^2 \!+\! a_1(n) K \sigma_x}.
    \label{eqn:Tradeoff2}
\end{align}
\end{subequations}
\end{theorem}

\begin{IEEEproof}
The theorem follows by noting that  $\zeta(T_{\rm s})$ in \eqref{57} is monotonically decreasing in $T_{\rm s}$, while $E_{\rm ratio}$ in \eqref{eq:eratio_WSS} is a monotonically increasing function of $T_{\rm s}$. Consequently, both \eqref{58} and \eqref{59} are obtained by identifying the minimal/maximal value of $T_{\rm s}$ for which the constraint holds with equality, hence proving the theorem.
\end{IEEEproof}

In the following subsection we provide a few examples of energy-fidelity tradeoffs which arise from the above analysis.

\subsection{Examples}
\label{subsec:Examples}
\begin{figure}
	\centering
	\includegraphics[width=\figWidth]{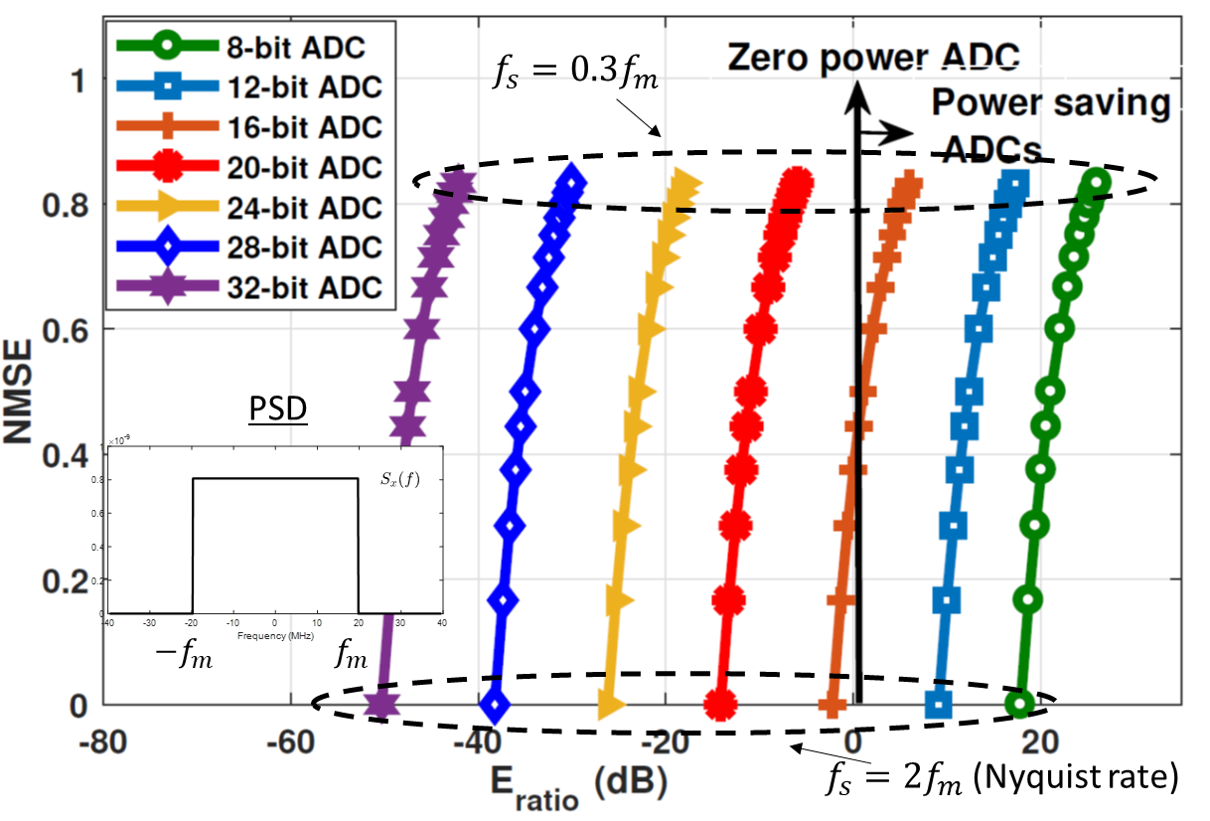}
	\caption{NMSE ($\zeta$) versus $E_{\rm ratio}$,  flat PSD.}
	\label{fig:flat}
\end{figure}

The characterization of the energy-fidelity tradeoff in Theorem \ref{thm:Tradeoffs} identifies the achievable energy ratio for a given recovery accuracy and vice versa. 
It also reveals the achievable energy ratio when \textit{e}Sampling a bandlimited signal of maximum frequency $f_m \geq 0$ with zero reconstruction error. In particular, combining Corollary \ref{cor:NyqSamp} and Theorem \ref{thm:Tradeoffs} indicates that this energy ratio is given by 
\begin{align}
\!\!E_{\rm ratio}\opt\!(0) \!= \!    \frac{\frac{\eta}{R_{\rm h}} (\frac{1}{2f_{\rm m}}-T_{\rm aq})\sigma_x^2}{a_2(n) K^2\sigma_x^2 \!+\! a_1(n) K \sigma_x}.
\label{eq:dis:1}
\end{align}
An example of how Theorem \ref{thm:Tradeoffs} is computed for arbitrary sampling rates is given in the following:
\begin{example}[Flat PSD]
	\label{exm:FlatTradeoff}
	Consider again the bandlimited signal with flat PSD of Example \ref{exm:Flat}. In this case, by \eqref{eq:nmse_flat_psd}, an NMSE of $\zeta(1/f_{\rm s})\leq\epsilon$ is guaranteed by using $f_{\rm s} \ge 2f_{\rm m}(1-\epsilon)$. Consequently, by Theorem \ref{thm:Tradeoffs}    the energy ratio under fidelity constraint $\epsilon$ for such signals is given by
	\begin{align}
	\!\!E_{\rm ratio}\opt\!(\epsilon) \!= \!    \frac{\frac{\eta}{R_{\rm h}} (\frac{1}{2f_{\rm m}(1-\epsilon)}-T_{\rm aq})\sigma_x^2}{a_2(n) K^2\sigma_x^2 \!+\! a_1(n) K \sigma_x}.
	\label{eq:dis:2}
	\end{align}
	The resulting energy-fidelity tradeoff curve for different numbers of quantization bits is depicted in Fig. \ref{fig:flat} under the following settings:  We use $K^2=20$, guaranteeing a probability of over 95\% that $|x(t)|\leq V_{\rm ref}$, while the ADC circuit parameters are set to $f_m=19.8$ MHz, $T_{\rm aq}=2.5$ ns, $C_{\rm u}=10$ fF, $C_{\rm c}=5$ fF, $C_{\rm s}=0.7$ fF, $R_{\rm h}=23.75~\Omega$, $A_{\rm k}=1.8$, $V_{\rm e}=0.05$ V, $\alpha_{\tau}=5$, $V_{\rm ref}=0.8$ V, $g=0.4$, and $\eta = 0.7$.  Finally, the signal power $\sigma_x^2$ is accordingly set to $\frac{V_{\rm ref}^2}{K^2}$.
\end{example}

The specific design parameters used in evaluating Fig. \ref{fig:flat} correspond to the \textit{e}Sampling ADC circuit design presented in Section \ref{sec:circuit_level_implementation}, and are in the typical ranges provided in previous works on ADC circuitry, e.g.,   \cite{6043594,4588351,Craninckx2007A60}. The efficiency of the energy harvesting system $\eta$ is in line with similar values reported for energy harvesting circuits in \cite{5117948, 5599946,7792628}. 

As expected, the achievable energy ratio in Example \ref{exm:FlatTradeoff} coincides with \eqref{eq:dis:1} when perfect recovery is required, i.e., $\epsilon = 0$.
The energy ratio characterized in \eqref{eq:dis:2}  is increased by reducing the sampling rate, which in turn increases the reconstruction error, $\epsilon$, as illustrated in Fig. \ref{fig:drawing}. 
The fundamental balance between these measures follows from the structure of \textit{e}Sampling ADCs, in which increasing the hold time degrades the ability to recover the signal from its samples, while allowing to harvest more energy. This unique property of \textit{e}Sampling can lead to ADCs which harvest more power than they consume, as  observed in Fig. \ref{fig:flat}. 

The results shown in Fig. \ref{fig:flat} demonstrate that an \textit{e}Sampling ADC with up to $12$ bits acquiring a bandlimited signal  can harvest more power than it consumes while sampling at Nyquist condition, and hence achieving zero-approaching reconstruction error. 
While the ability of \textit{e}Sampling ADCs to sample at Nyquist rate and zero-power is observed in Fig. \ref{fig:flat} for signals with flat PSDs, it  holds for arbitrary PSD shapes as long it is bandlimited to $f_{\rm m}$ and the variance of the signal is $\sigma_x^2$. This follows since by \eqref{eq:eratio_WSS}, the energy ratio for a given sampling rate and signal variance does not depend on the shape of the PSD.
 However, for the ADC to operate at zero power with higher resolution quantization, one has to sample below the Nyquist rate and hence compromise in reconstruction error. In particular, each of the curves in Fig. \ref{fig:flat} reaches zero NMSE for $f_{\rm s} = 2 f_{\rm m}$, while reducing the sampling rate allows achieving improved energy ratio at the cost of reduced reconstruction accuracy, reaching poor recovery performance of $\zeta = 0.85$ as $f_{\rm s}$ is reduced to   $0.3 f_{\rm m}$. It is emphasized that for a given sampling rate, \textit{e}Sampling ADCs implement the same acquisition mapping as conventional S/H ADCs, and thus their the ability to harvest energy using \textit{e}Sampling ADCs does not come at the expense of conversion accuracy. However, \textit{e}Sampling provides to possibility to increase the amount of energy harvested by increasing the sampling interval, which in turn may degrade the ability to recover the analog signal.

\begin{figure}
	\centering
	\includegraphics[width=\figWidth]{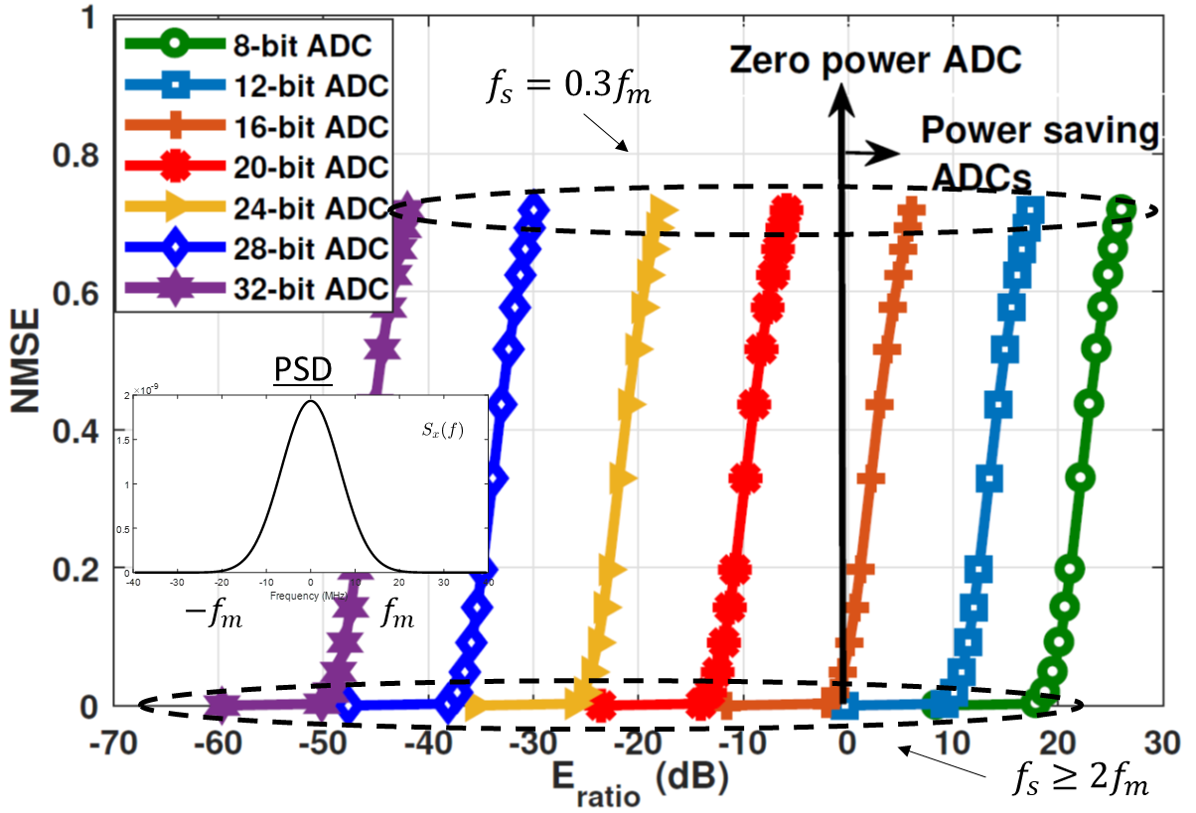}
	\caption{NMSE ($\zeta$) versus $E_{\rm ratio}$, 
		unimodal PSD. }
	\label{fig:unimodal}
\end{figure}

 
	As discussed above, while the recovery NMSE  depends not only on the sampling rate but also on the shape of the PSD $S_x(f)$ \eqref{57}, the energy ratio for a fixed sampling rate is affected only by the overall input energy $\sigma_x^2 = \int S_x(f) df$ \eqref{eq:eratio_WSS}. This follows from the fundamental difference between the two objectives of \textit{e}Sampling, i.e., acquisition and energy harvesting: The purpose of acquisition is to allow the complete signal, whose profile depends on the shape of its PSD, to be recovered from its digital representation. However, energy harvesting aims at extracting energy from the signal without having to maintain sufficiency or to avoid distorting the signal, and is invariant of the specific shape of its PSD. The dependency of the energy-fidelity tradeoff on the PSD profile is demonstrated in the following two examples which, unlike Example \ref{exm:FlatTradeoff}, consider non-purely-bandlimited signals:

\begin{example}[Unimodal PSD]
	\label{exm:UnimodalTradeoff}
	Let $x(t)$ be a WSS signal with a PSD given by $S_x(f)=\alpha e^{-\frac{f^2}{2\sigma^2}}$, where $\alpha=\frac{\sigma_x^2}{\sqrt{2\pi\sigma^2}}$ such that $\int_{-\infty}^{\infty}S_x(f)df=\sigma_x^2$. The parameter $\sigma^2$ controls the PSD width, and is set to $\sigma = f_{\rm m}/3$. 
	The resulting energy-fidelity tradeoff computed via Theorem \ref{thm:Tradeoffs} under the ADC circuit parameters used in Example \ref{exm:FlatTradeoff} is depicted in Fig. \ref{fig:unimodal}, along with an illustration of the PSD. 
\end{example}

\begin{figure}
	\centering
	\includegraphics[width=\figWidth]{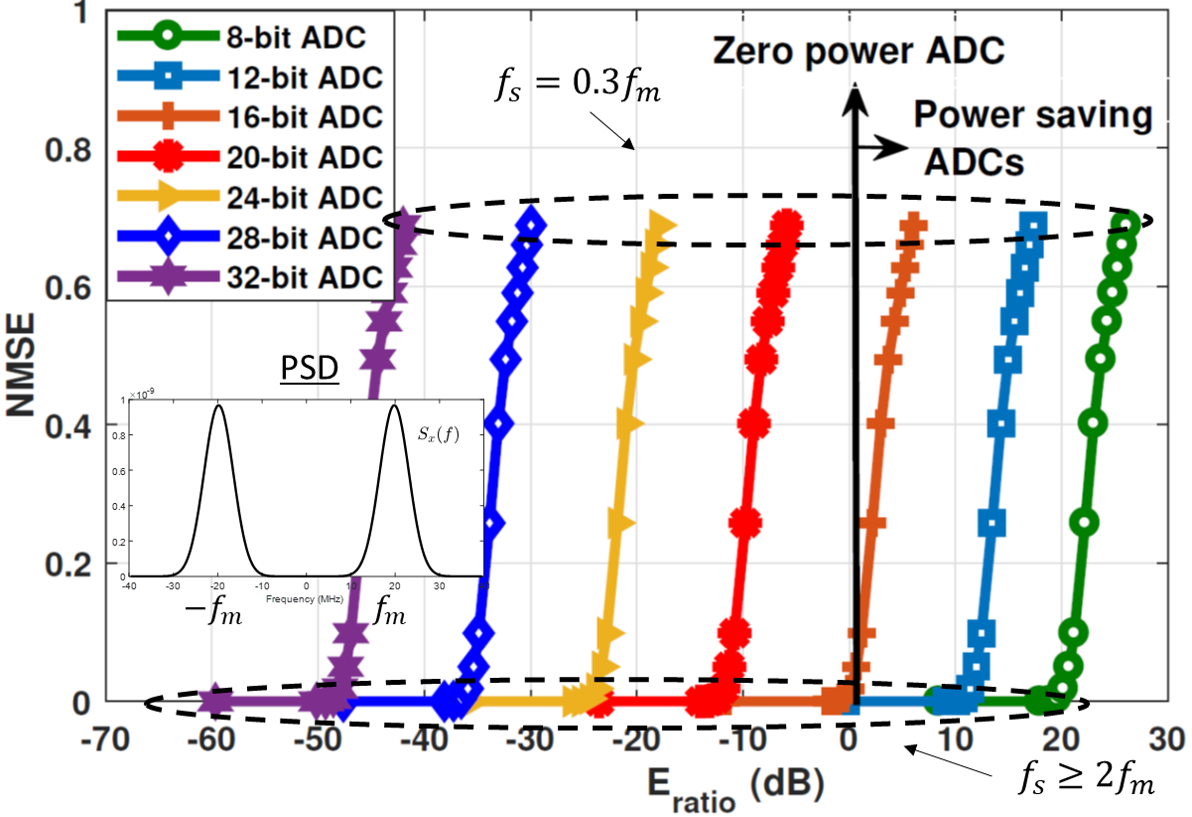}
	\caption{NMSE ($\zeta$) versus $E_{\rm ratio}$,  multimodal PSD.}
	\label{fig:multimodal}
\end{figure}

\begin{example}[Multimodal PSD]
	\label{exm:MultimodalTradeoff}
	Let $x(t)$ be a WSS signal with a PSD $S_x(f)=\frac{\alpha}{2}(e^{-\frac{(f-f_{\rm m})^2}{2\sigma^2}}+e^{-\frac{(f+f_{\rm m})^2}{2\sigma^2}})$. Here,  $\sigma$ is set to  $\sigma = f_{\rm m}/6$. This PSD profile and the  energy-fidelity tradeoff evaluated using Theorem \ref{thm:Tradeoffs} under the ADC circuit parameters used in Example \ref{exm:FlatTradeoff} is depicted in Fig. \ref{fig:multimodal}. 
\end{example}

 These examples illustrated in Figs. \ref{fig:unimodal} and \ref{fig:multimodal} demonstrate that \textit{e}Sampling ADCs applied to signals with such spectral profiles can operate with zero power for up to $n=16$ bits of quantization resolution, while achieving approximately ideal reconstruction. 
 Observing Figs. \ref{fig:unimodal}-\ref{fig:multimodal} and comparing them to Fig. \ref{fig:flat}, we note that different PSD profiles lead to different energy-fidelity curves. This property is solely due to the dependence of the achievable NMSE on the PSD, which follows from Theorem \ref{thm:MSE}, since   both the amount of energy harvested from a stationary signal as well as that consumed in  \textit{e}Sampling do not depend on the spectral profile of the signal, but on the sampling rate and the  variance $\sigma_x^2$. 
 
 In particular, the amount of energy harvested \eqref{expected_energy_harvested} when  \textit{e}Sampling at $f_{\rm s} = 2f_{\rm m}$ is numerically evaluated as $0.35$ pJ, while the corresponding amount of energy consumed \eqref{eq:ehold} when using $n=8$ bit quantizers is  $21.2$ pJ. 
%
This implies that the \textit{e}Sampling ADC is able to harvest much more energy from the signal than it consumes in converting it into a digital representation, as the energy ratio indicates an energy gain of $17.8$ dB. In particular, it is observed that \textit{e}Sampling ADCs operating with up tp $16$ bits per sample are capable of saving power. However, this mode of operation comes at the cost of increased NMSE for higher values of $n$. The examples presented in this subsection indicate that the power consumption of high resolution ADCs can be notably reduced and even mitigated by properly combining acquisition and energy harvesting via \textit{e}Sampling. In Section~\ref{sec:circuit_level_implementation} we demonstrate that these results do not follow only from a numerical evaluation of our theoretical results, but also reflect the performance in terms of recovery accuracy and energy efficiency of a dedicated \textit{e}Sampling ADC circuit design.

\subsection{Discussion}
\label{subsec:discussion}
Our characterization in the previous subsections focuses on the general family of stationary signals. When the signal obeys some structure, e.g., it is known to be sparse in the frequency domain, ideal recovery can be achieved at low sampling rates using generalized sampling methods \cite{eldar2015sampling}, allowing to harvest more energy without affecting the recovery NMSE. This indicates that the energy-fidelity tradeoff of \textit{e}Sampling ADCs can be further improved by accounting for structured signals, as commonly encountered in communication \cite{cohen2018analog} and radar \cite{cohen2018sub} systems.  We leave the analysis of \textit{e}Sampling of structured signals for future work.

The fact that 
\textit{e}Sampling gives rise to ADCs which operate with zero power and can even harvest more energy than they consume, makes it an attractive technology for low-power systems, such as internet of things devices, sensor networks, as well as wearable and implantable medical units. However, the applicability of the proposed \textit{e}Sampling ADC is limited in some scenarios since its architecture is based on S/H ADCs. For example, S/H ADCs typically operate at sampling rates below $1$ GHz, and are not suitable for operating at extremely high sampling rates, where flash ADCs are more commonly used. While we conjecture that the concept of \textit{e}Sampling, namely, the integration of energy harvesting into signal acquisition, can also be combined with alternative ADC technologies other than S/H, we leave this study for future research.
 
While our analysis focuses on WSS signals for analytical tractability,  the proposed \textit{e}Sampling  ADCs applies to a much broader family of acquired analog signals. For example the \textit{e}Sampling  ADC circuitry detailed in the following section is experimented  when acquiring a sinusoidal signal, demonstrating its ability to accurately reconstruct the signal in a power saving manner. 
Furthermore, our proposed analysis is based on linear recovery, being a common reconstruction framework in sampling theory. In particular, the reconstruction of Nyquist rate sampled bandlimited signals, shift-invariant signals, and various other structures studied in the literature, is based on linear filtering \cite{eldar2015sampling}. However, the architecture of the \textit{e}Sampling  ADC is invariant to the reconstruction mechanism, and alternative recovery schemes would result in a different characterization of the energy-fidelity tradeoff.

\color{black}

\section{\textit{e}Sampling ADC Circuit-level Design}
\label{sec:circuit_level_implementation}
In order to demonstrate the hardware feasibility of the concept of \textit{e}Sampling, we present the circuit-level design of such a device. In particular, we design an \textit{e}Sampling ADC circuit based on the model shown in Fig.~\ref{fig:eSampling_system_model} using standard  65~nm  CMOS technology, and carry out its experimental study using a Cadence Virtuoso platform. 
In order to design the \textit{e}Sampling ADC based on the high-level architecture illustrated in Fig. \ref{fig:eSampling_system_model}, one has to design its three main sub-blocks: The two-way switch  $\tilde{S}$; the quantizer logic; and the energy harvesting circuit. We thus first elaborate on each of these sub-blocks, after which we present the experimental study.
\begin{figure}
\centering
\includegraphics[width=\columnwidth]{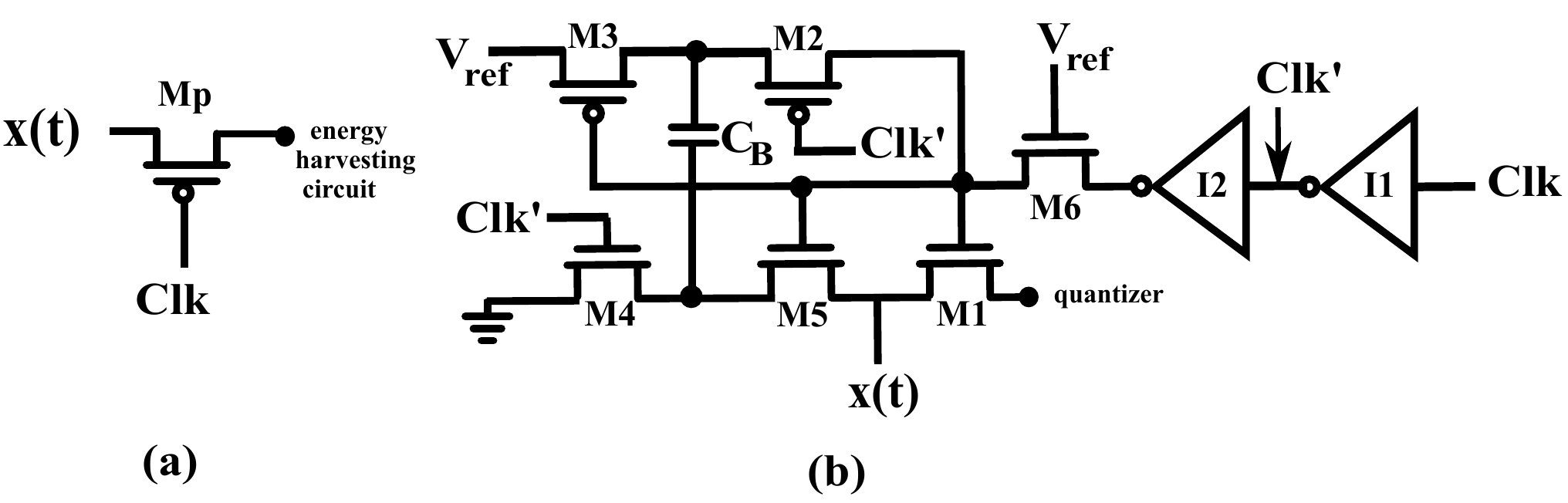}
\caption{Circuit diagram of (a) PMOS transistor switch, (b) NMOS bootstrapped switch.}
\label{switch}
\end{figure}
\subsection{Two-way switch}
The two-way switch $\tilde{S}$ allows the input signal to be connected to the hold capacitor during acquisition phase and to the energy harvesting circuit during the hold phase. In our design, $\tilde{S}$ is implemented\footnote{The term `\textit{implement}' used here implies the design/simulation of the circuit in Cadence Virtuoso platform, in line with the similar usage of this terminology in  \cite{liu201010,6936944,8727467,8676062, 8017456}.} using two one-way switches, one for each operation phase, namely, when one switch is open, the other is closed. Each of these switches is realized using a different topology. The switch designed to connect the input signal to the energy harvesting circuit is implemented using a PMOS transistor, as illustrated in Fig. \ref{switch}(a). The PMOS transistor turns ON when the clock signal $Clk$ is at logic '0', indicating that hold phase is active. When $Clk$ is at logic '1', it turns OFF and isolates the input signal from the next block. In order to allow both switches of $\tilde{S}$ to utilize the same single clock pulse, the switch designed to connect the input signal with the quantizer is implemented using an NMOS transistor, which turns on when $Clk$ is at '1'. 

The on-resistance of a MOS transistor, which determines the value of $R_{\rm on}$ in \eqref{eq:Taq}, is sensitive to fluctuations in the input signal and may vary accordingly \cite{razavi1995principles}. 
%
Such variations in $R_{\rm on}$ may introduce a non-linear distortion at the output of the ADC. To avoid such distortion, we use an NMOS bootstrap switch to connect the input signal to the quantizer, which ensures a constant $R_{\rm on}$, as proposed in \cite{7258484}. The design of the NMOS transistor based bootstrapped switch used in this work is illustrated in Fig. \ref{switch}(b). To achieve nearly constant $R_{\rm on}$, the gate of the transistor $M1$ in Fig. \ref{switch}(b) is bootstrapped using two PMOS transistors $M2$ and $M3$, three NMOS transistors $M4,~M5$ and $M6$, and one capacitor $C_{\rm B}$, following \cite{7258484}. Two CMOS inverters $I1$ and $I2$ are also employed in the structure to generate the required clock signals needed for proper operation of the switch.

The value of the on-resistance  $R_{\rm on}$ as well as the hold capacitor $C_{\rm h}$ affect the setting of the acquisition time $T_{\rm aq}$, as follows from \eqref{eq:Taq}. To maximize the amount of energy harvested, small values of $T_{\rm aq}$ are preferable, so that more time could be allocated to harvesting the input signal energy. Reducing $R_{\rm on}$ requires increasing the width of the transistors \cite{razavi1995principles}, which in turn increases the device capacitance, and thus reduces the operating speed of the ADC. In addition, wider devices may result in charge injection \cite{chen1995weak}, which degrades the signal-to-noise-distortion ratio (SNDR) of the ADC, and hence the performance of the ADC. Alternatively, employing small values for $C_{\rm h}$ results in mismatch issues and sampling noise, which degrade the ADC conversion accuracy \cite{7890,chen2014capacitor}. These drawbacks require the acquisition time $T_{\rm aq}$ to be large enough such that the ADC performance is not compromised, and is in fact the primary reason S/H ADCs are typically limited to operate with sampling rates below 1 GHz, as discussed in Subsection \ref{subsec:discussion}. 

\subsection{Quantizer}
\begin{figure}
\centering
\includegraphics[width=\figWidth]{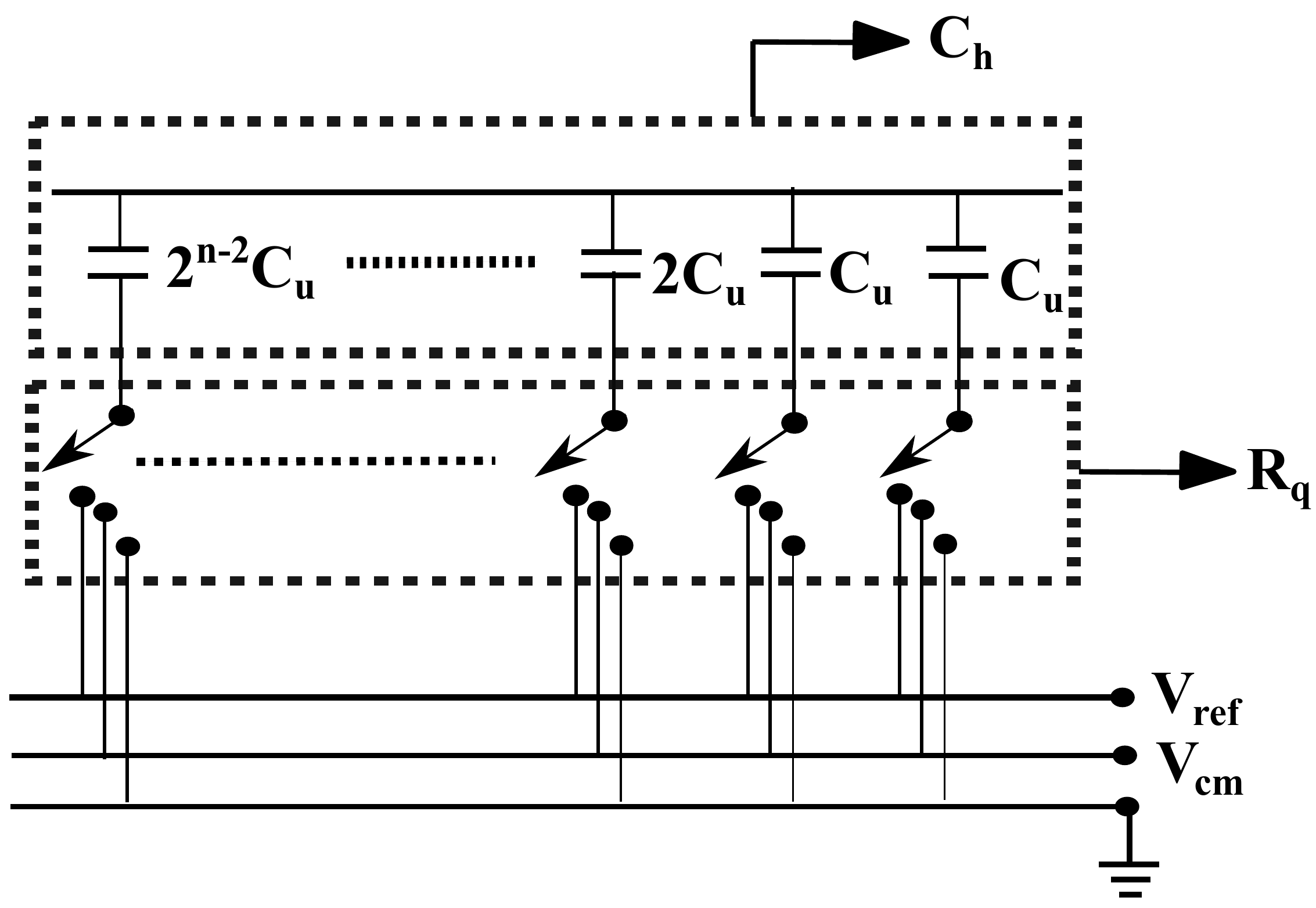}
\caption{DAC capacitor array schematic diagram.} 
\label{fig:dac_cap_array}
\end{figure}

The dedicated  \textit{e}Sampling ADC circuit design is based on S/H SAR ADC architectures \cite{6341363,6043594,hariprasath2010merged} as illustrated in Figs. \ref{fig:1}-\ref{fig:eSampling_system_model}. Such quantizers generally consist of a DAC, a voltage comparator and a SAR logic, which map the voltage of the hold capacitor (also known as the total capacitance of DAC array) into an $n$-bit value by successively refining the digital representation using a binary search algorithm. In our \textit{e}Sampling ADC circuit we use a single-ended merge capacitor switching (MCS) based SAR ADC. For such devices, the total capacitance of the DAC array is $C_{\rm h}=2^{n-1}C_{\rm u}$, where $C_{\rm u}$ is the unit capacitance of the DAC array, as illustrated in Fig.~\ref{fig:dac_cap_array}.

In particular, during acquisition phase the input signal $x(t)$ is connected to the top plate of the DAC capacitor array, while the bottom plate is connected to the common mode voltage, i.e., $V_{\rm cm}=\frac{V_{\rm ref}}{2}$. Once the acquisition phase is over, the voltage at the top plate of the DAC capacitor array is reduced by common mode voltage, and hence equals to $x(kT_s)-V_{\rm ref}/2$. The top plate of the DAC capacitor array is connected to the positive terminal of the comparator, while the negative terminal of the comparator is grounded.
The comparator then compares the voltage of its positive terminal with its negative terminal. If the voltage at the positive terminal is higher than the negative terminal, the comparator yields an output of logic `1', else logic `0'. The output of the comparator is passed to the SAR logic, which resolves the most significant bit (MSB). The decision on the MSB is fed back to the DAC and the bottom plate of the largest capacitor of DAC capacitor array is switched from $V_{\rm cm}$ to ground (if MSB=1) or $V_{\rm ref}$ (if MSB=0). This operation changes the voltage at the top plate of the DAC capacitor array, and a new decision is made by the comparator, which is sent to the SAR logic to resolve the second MSB and so on. The process continues for all $n$ bits. The overall resistance of the switches is determined by the binary scale switch resistance, $R_{\rm q}$, as illustrated in Fig. \ref{fig:dac_cap_array}.

As discussed in Subsection \ref{subsec:SH}, the energy consumption of S/H SAR ADCs is effectively dictated by its quantization sub-blocks. Therefore in the following, we detail the circuitry used for the quantizer along with its energy usage per sample.

%
 
The voltage comparator is implemented using a dynamic latch. The energy consumed per sample of a dynamic latch comparator is given by \cite{6043594}
\begin{align}
    E_{\rm c}=nC_{\rm c}V_{\rm ref}^2+  2V_{\rm ref} \gamma_{\rm n}, \label{eq:energy_comparator}
\end{align}
where $\gamma_{\rm n}:=V_{e}C_{\rm c} \big(n\ln{1/A_{\rm k}}+\frac{n(n+1)}{2}\ln{2}+n \big)$, $C_{\rm c}$ is the capacitive load of the comparator, $A_{\rm k}$ is the gain during regenerative phase, and $V_{\rm e}$ is the ratio of the drain current of the device with its trans-conductance \cite{4588351}.
The SAR logic is realized using two arrays of shift registers that operate in serial-in-parallel-out and parallel-in-parallel-out modes \cite{5771068}. Each register is implemented using a D flip-flop circuit, and the resulting energy consumption is given by \cite{6043594}
\begin{equation}
    E_{\rm sl}=16n^2g C_{\rm s}V_{\rm ref}^2, \label{eq:energy_SAR_logic}
\end{equation}
 where $C_{\rm s}$ is the input capacitance of the D flip-flop, and $g \in [0,1] $ is the total activity parameter of the SAR logic. 
Finally, the DAC is based on a binary-weighted capacitive DAC,  designed using the MCS scheme \cite{hariprasath2010merged}. 
The energy consumption of the MCS DAC  is given by \cite{hariprasath2010merged}
\begin{align}
    E_{\rm DAC}&=\rho_{\rm n} n C_{\rm u}V^2_{\rm ref},
    \label{EDAC}
\end{align}
where $\rho_{\rm n}=\sum_{i=1}^{n-1}2^{n-3-2i}(2^i-1)$.

To summarize, the total energy consumption during hold phase of our dedicated \textit{e}Sampling ADC circuit design, which dictates the overall energy consumed per sample, is given by
\begin{align}
    E_{\rm hold}&= E_{\rm DAC} +E_{\rm sl} + E_{\rm c} \notag \\
    &\stackrel{(a)}{=} V^2_{\rm ref}\left(  \rho_{\rm n} C_{\rm u}+nC_{\rm c}+ 16n^2 C_{\rm s}g \right)+2V_{\rm ref} \gamma_{\rm n}, \label{eq:ehold}
\end{align}
where $(a)$ follows from \eqref{eq:energy_comparator}, \eqref{eq:energy_SAR_logic}, and \eqref{EDAC}. The energy term in \eqref{eq:ehold} obeys the second-order polynomial model of \eqref{53},  used in our analysis of \textit{e}Sampling ADCs in Section \ref{sec:Analysis}. 
 


\subsection{Energy Harvesting Circuit}
\label{subsec:EnregyCircuit}
The proposed \textit{e}Sampling ADC harvests the input signal energy during hold phase and stores this energy in a capacitor, $C_{\rm EH}$. As detailed in Subsection \ref{subsec:eSampling}, energy harvesting circuits typically consist of a capacitor, in which the harvested energy is stored, and a signal conditioning circuit, whose purpose is to facilitate the charging of the capacitor. In our design, we do not include a signal conditioning circuit and forward the input signal directly to $C_{\rm EH}$ during hold time. This simplified design is sufficient for our experimental purposes, where we use synthetic controlled input signals with strictly positive voltage values. However, in order to achieve efficient energy harvesting of a low voltage complex rapidly alternating signals, one should also include signal conditioning devices, such as a rectifier, voltage regulator, and DC-DC converter. 

 To quantify the maximum amount of energy that can be harvested in an analytically tractable manner, we consider the case where the input signal is approximately constant during the hold phase, i.e., $x(t) \approx x(T_s)$ for each $t \in [T_{\rm aq}, T_{\rm s}]$. The purpose of this approximation is to facilitate characterizing the amount of energy harvested in a tractable manner. In addition, we focus on the scenario in which the capacitor is empty at the beginning at the hold phase, namely, the voltage on the capacitor $C_{\rm EH}$, denoted $ V_{\rm EH}(0)$, satisfies $V_{\rm EH}(T_{\rm aq}) = 0$. In this setup, the capacitor voltage at the end of the hold phase, i.e.,  at time instance $t=T_{\rm s}$, is given by
\begin{align}
    V_{\rm EH}(T_{\rm s})  \approx x(T_s)\left(1-e^{-\frac{T_{\rm h}}{R_{\rm h}C_{\rm EH}}}\right), 
    \label{eq:actual_voltage_harvested}
\end{align}
where, as defined in Subsection \ref{subsec:Problem}, $R_{\rm h}$ is the resistance of the energy harvesting circuit. This resistance is dictated here by the on-resistance of the PMOS transistor in the two-way switch.
The amount of energy harvested in such a sampling interval is given by
\begin{align}
    E_{\rm h} &= \frac{1}{2}C_{\rm EH}V_{\rm EH}^2(T_{\rm s}) \nonumber \\
  & \stackrel{(a)}{\approx}  \frac{1}{2}C_{\rm EH}\left(1-e^{-\frac{T_{\rm h}}{R_{\rm h}C_{\rm EH}}}\right)^2 x^2(T_s) \nonumber \\
   & \stackrel{(b)}{\approx}  \frac{1}{2T_{\rm h}}C_{\rm EH}\left(1-e^{-\frac{T_{\rm h}}{R_{\rm h}C_{\rm EH}}}\right)^2 \int\limits_{T_{\rm aq}}^{T_{\rm h}} |x(t)|^2dt,
    \label{eq:actual_energy_harvested}
\end{align}
where $(a)$ follows from \eqref{eq:actual_voltage_harvested}, and $(b)$ stems from the fact that the input is approximately constant during the hold phase. Comparing \eqref{eq:actual_energy_harvested} and \eqref{expected_energy_harvested} reveals that the efficiency of this simple energy harvesting circuit can be approximated as 
\begin{equation}
    \eta \approx \frac{R_{\rm h} C_{\rm EH}}{2T_{\rm h}}\left(1-e^{-\frac{T_{\rm h}}{R_{\rm h}C_{\rm EH}}}\right)^2.
    \label{eqn:EHeta}
\end{equation}
 
 The expression for the energy harvesting efficiency in \eqref{eqn:EHeta} can be used to provide guidelines for determining the capacitance $C_{\rm EH}$ used in the circuit. In particular, it can be shown that \eqref{eqn:EHeta} is maximized when $C_{\rm EH} \approx 0.796\frac{T_{\rm h}}{R_{\rm h}}$. However, the derivation of \eqref{eqn:EHeta} is carried out assuming that the capacitor is empty at the beginning of the hold phase. This implies that its stored energy is transferred to some external storage device, e.g., battery, after each sample. In practice, energy transfer typically takes much longer than a single sampling interval, and thus it is preferable to carry out such a transfer only once every multiple samples. This is achieved by using a capacitor with a larger value of $C_{\rm EH}$, which allows to store more energy and provides a nearly constant voltage at the load, but requires more time to charge. In particular, in our experimental setup detailed in Subsection \ref{subsec:SimCircuit}, we set
 $C_{\rm EH} = 42.2\frac{T_{\rm h}}{R_{\rm h}}$, which  results in the capacitor taking approximately $340$ samples to charge up. Under such a setting,  the period dedicated to transferring its energy once it is fully charged, during which energy harvesting is inactive, has only a minor effect on the overall harvested energy.

\subsection{Experimental study}
\label{subsec:SimCircuit}
To validate that the energy saving potential of \textit{e}Sampling ADC observed in Section \ref{sec:Analysis} also reflects its behavior in a real world environment, we next evaluate the \textit{e}Sampler circuit design. To that aim, a schematic of the \textit{e}Sampling ADC circuit has been created in Cadence Virtuoso platform based on the circuit-level design detailed in the previous subsections. The proposed \textit{e}Sampling ADC operates at a sampling frequency of $40$ MHz with an $n = 8$ bit quantizer. For our experimental purpose, we use a sinusoidal signal, being a common benchmark for evaluating the accuracy of ADC circuits \cite[Ch. 2]{kester2005data}. The maximum frequency of the input signal is $19.8$ MHz, thus the sampling rate satisfies the Nyquist condition. The amplitude of the signal varies from 0 to $V_{\rm ref}$. Here, we use an energy harvesting capacitor of  $C_{\rm EH}=40$ nF, while the remaining  parameters are the same those detailed in Examples \ref{exm:FlatTradeoff}-\ref{exm:MultimodalTradeoff}.

\begin{figure}
\centering
\includegraphics[width=\figWidth, height=2.7in]{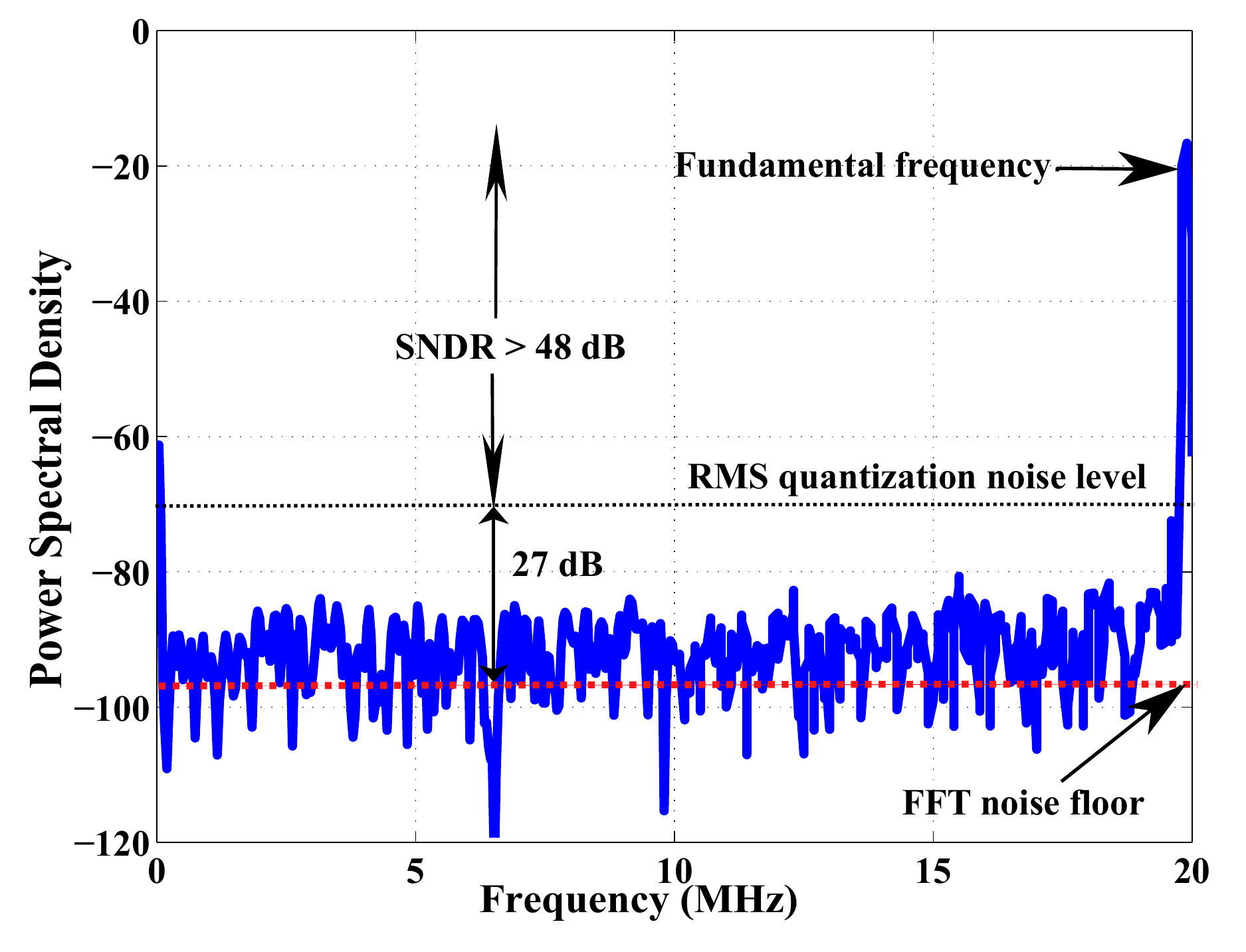}
\caption{FFT plot of reconstructed signal for 8 bit \textit{e}Sampling ADC.} 
\label{fig:fft}
\end{figure}

We first assert that the \textit{e}Sampling ADC is indeed capable of accurately reconstructing the signal sampled at the Nyquist rate. To that aim, we depict the  fast Fourier transform (FFT) of the reconstructed signal, computed using a 1024-point FFT, in Fig.~\ref{fig:fft}. As expected, the FFT noise floor is determined by the SNDR due to quantization, computed by the $6$ dB rule of thumb as approximately $48$ dB, with the additional FFT processing gain of  $10\log_{10}(1024/2) \approx 27$ dB \cite[Ch. 2]{kester2005data}. In particular, the gap between the noise floor observed in Fig.~\ref{fig:fft} and the energy of the signal at its central frequency of $19.8$ MHz, is roughly $75.52$ dB, settling with the theoretical performance of ADCs satisfying Nyquist condition, and indicating that the designed \textit{e}Sampling ADC accurately reconstructs the observed analog signal. 


\begin{figure}
	\centering
\includegraphics[width=\figWidth, height=2.7in]{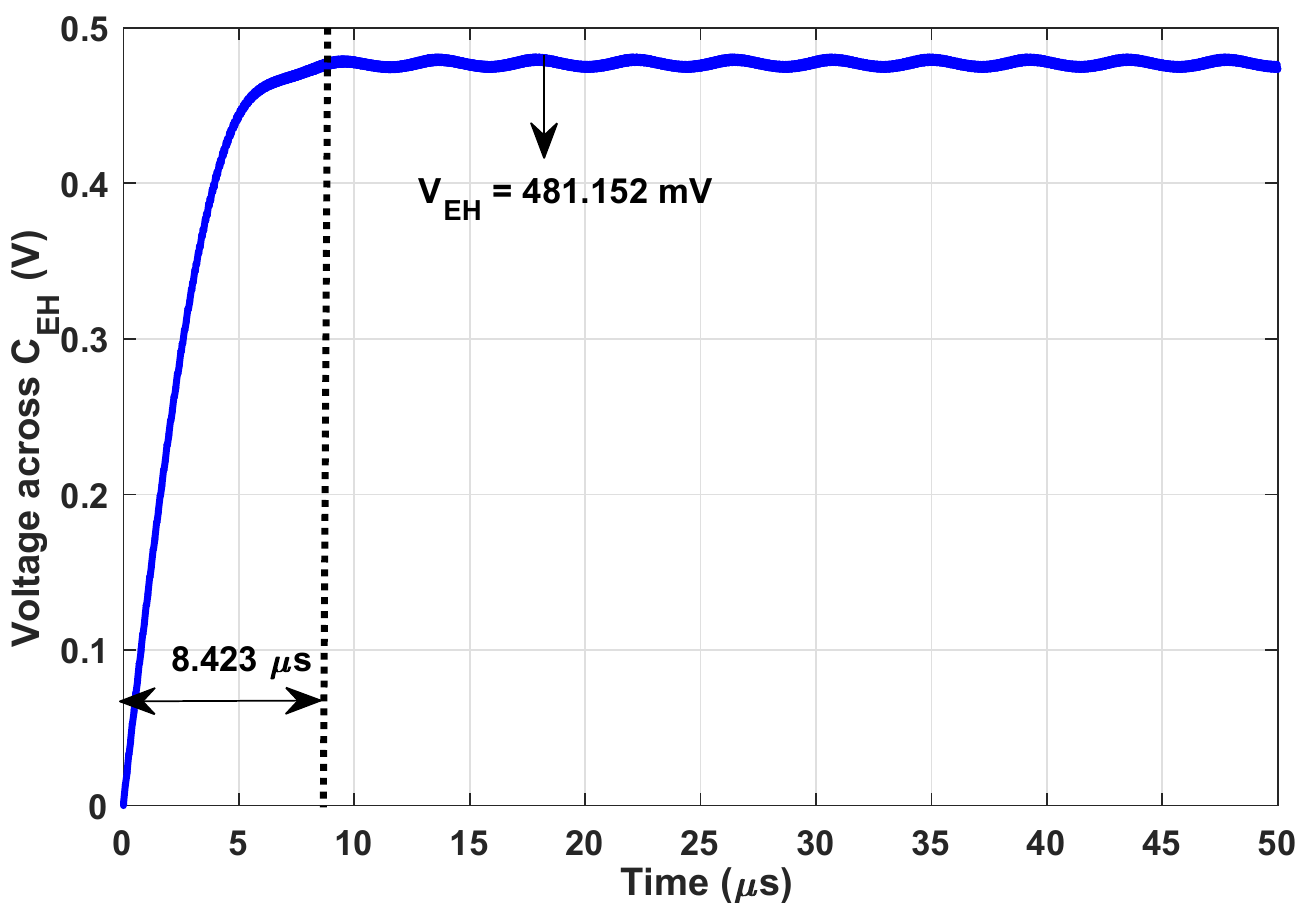}
\caption{Voltage obtained across $C_{\rm EH}$ for 8 bit \textit{e}Sampling ADC.}
\label{fig:V_C_EH}
\end{figure}

Next, we focus on the energy harvesting circuit of the designed \textit{e}Sampler, in order to identify how many sampling rounds are required for the capacitor to charge up.  
To that aim, we plot in Fig.~\ref{fig:V_C_EH} the voltage on the energy harvesting capacitor over time. Observing Fig.~\ref{fig:V_C_EH}, we note that for the given input signal, the capacitor reaches a steady level of  $V_{\rm EH}=481.152$ mV after  $8.432$ $\mu$s, which correspond to $337$ samples at $40$ MHz. Based on Fig.~\ref{fig:V_C_EH}, we design the \textit{e}Sampling ADC to transfer the energy stored in its energy harvesting capacitor once every $337$ samples. We dedicate approximately $1.5$ $\mu$s for each transfer, during which the energy harvesting circuit is inactive, resulting in each cycle of harvesting and transferring taking approximately $500$ samples. Consequently, the effective amount of energy harvested per sample of the  \textit{e}Sampling ADC is given by
\begin{equation}
    E_{\rm h} = \frac{1}{2\cdot 500}C_{\rm EH} V_{\rm EH}^2(337\cdot T_{\rm s}) = 9.26 pJ.
    \label{eqn:TrueEh}
\end{equation}


The amount of energy harvested per sample, evaluated in \eqref{eqn:TrueEh} based on the experiment in Fig. \ref{fig:V_C_EH}, does not represent the overall energy balance of the \textit{e}Sampling ADC, as it accounts only for the amount of energy harvested. Therefore, to demonstrate that the  \textit{e}Sampling ADC circuit design not only accurately recovers the signal and harvests energy, but also saves more energy than it consumes, we next evaluate both the energy harvested and the energy consumed by the ADC circuit. The average energy consumption of our designed circuit is computed by evaluating the current drawn from its reference source $V_{\rm ref}$, denoted $I_{\rm ref}(t)$, and thus the energy consumed at each time instance can be obtained by
\begin{equation}
    E_{\rm cons}(t)=V_{\rm ref}\int_{0}^{t}I_{\rm ref}(\tau)d\tau.
    \label{eqn:ECons}
\end{equation}

\begin{figure}
	\centering
\includegraphics[width=\figWidth, height=2.7in]{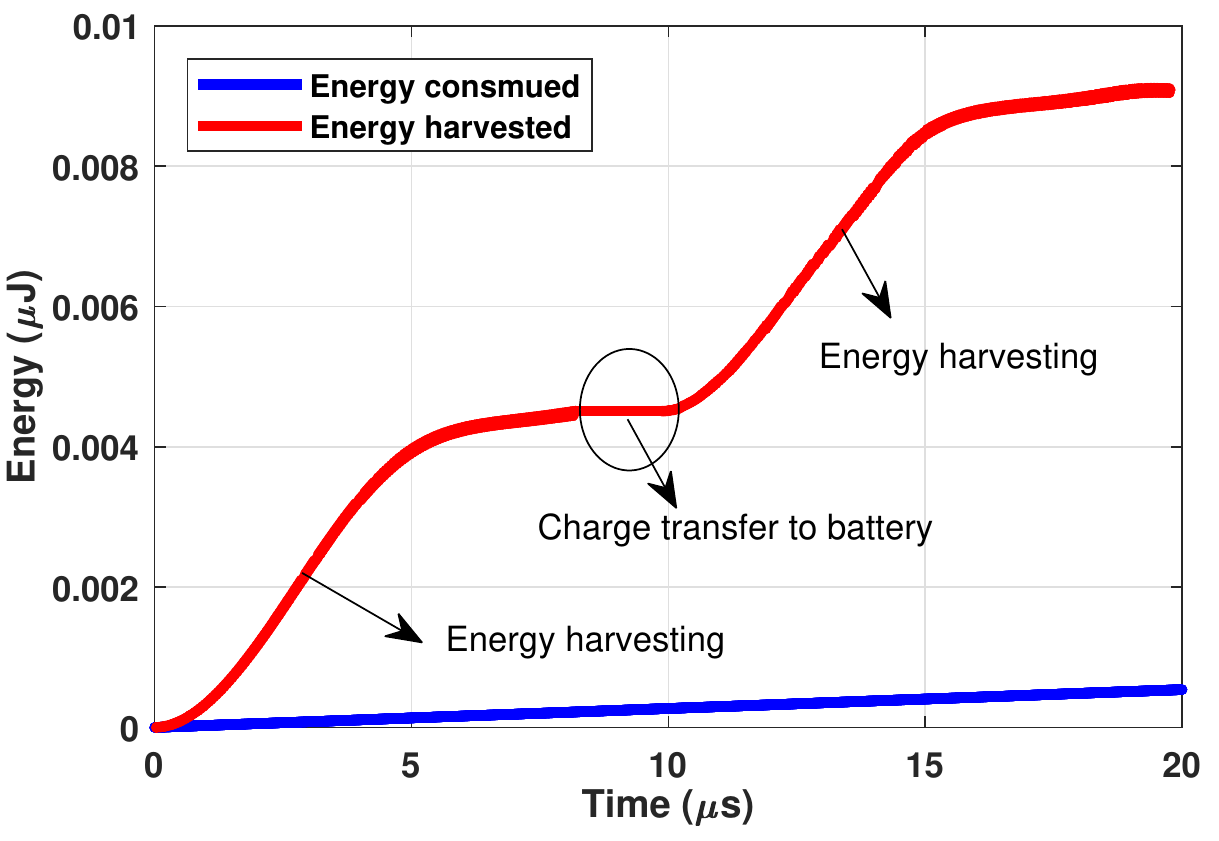}
\caption{Total energy harvested and energy consumed versus time for an 8 bit \textit{e}Sampling ADC.}
\label{fig:Eh_con_time_plot}
\end{figure}

The resulting growth of both the energy consumed and the energy harvested are depicted in Fig. \ref{fig:Eh_con_time_plot}. 
Observing Fig. \ref{fig:Eh_con_time_plot}, we note that the \textit{e}Sampling ADC harvests much more energy than it consumes, while still being able to accurately reconstruct its input signal as demonstrated in Fig. \ref{fig:fft}. In particular, the consumed energy is shown to grow in an approximately linear manner, with an average energy consumption of  $0.56$ pJ per sample. The maximal amount of energy which can be obtained is dictated by the external battery, to which the harvested power is periodically transferred. Comparing this to \eqref{eqn:TrueEh} reveals that the true energy ratio of the \textit{e}Sampling ADC, which periodically transfers its harvested energy to an external battery, is approximately $12.1$ dB, which is within a relatively small gap from the theoretical results observed in Subsection \ref{subsec:Examples}. This gap can be further reduced by using more advanced energy harvesting circuitry, compared to the simplistic design detailed in Subsection \ref{subsec:EnregyCircuit}. In particular, using more advanced harvesting architecture can increase the efficiency $\eta$, allowing to achieve improved energy-fidelity tradeoffs compared to those observed here. Nonetheless, despite its relatively simple architecture, the \textit{e}Sampling ADC circuit design is still shown to be able to achieve accurate reconstruction while harvesting substantially more energy than it consumes. 



\section{Conclusion}  
\label{sec:Conclusions}    
In this paper, we proposed the \textit{e}Sampling ADC architecture, which modifies the traditional conversion process of a S/H ADC to harvest energy from the discarded portion of the input signal. We analyzed the amount of energy which can be harvested from stationary signals and characterized the underlying fundamental tradeoff between energy harvested and reconstruction fidelity which arises from the joint acquisition and energy harvesting paradigm.  
Our theoretic characterization reveals that an \textit{e}Sampling ADC with up to 12 bits can harvest more power than it consumes, when sampling both  bandlimited signals and non-bandlimited ones at a sampling rate allowing recovery with negligible error.
Then, we presented a circuit-level design of an \textit{e}Sampling ADC using CMOS 65 nm technology demonstrating the feasibility of this concept. Our experimental results validated the theoretical observations, showing that an \textit{e}Sampling 8-bit ADC circuit applied to a sinusoidal signal harvests more power than it consumes while recovering the analog signal in a nearly perfect manner.  

     \bibliography{IEEEabrv,ref,sampling_EH}
 \bibliographystyle{IEEEtran}

\end{document}